\documentclass[sigconf]{acmart}

\usepackage{epsfig, endnotes, graphicx, xcolor, xspace, subcaption}
\usepackage[all]{nowidow}

\usepackage{multirow}
\usepackage{booktabs}
\usepackage{tabularx}
\usepackage{array}
\usepackage{arydshln}
\usepackage{balance}

\setcopyright{none} %
\acmConference[]{}{}{}
\acmBooktitle{}
\acmDOI{}
\acmISBN{}
\renewcommand\footnotetextcopyrightpermission[1]{} %

\newcommand{\eat}[1]{}

\newcommand{\nbmFull}{National Broadband Map\xspace}
\newcommand{\nbm}{NBM\xspace}
\newcommand{\bdcFull}{Broadband Data Collection\xspace}
\newcommand{\bdc}{BDC\xspace}

\begin{document}

\pagestyle{empty}

\title{\PaperTitle}
\title[Red is Sus: Automated Identification of Low-Quality Service Availability Claims in the US NBM]{Red is Sus: Automated Identification of Low-Quality Service Availability Claims in the US National Broadband Map}

\newcommand{\authorhspace}{\hspace{-0.2cm}}

\author{Syed Tauhidun Nabi}
\affiliation{%
  \institution{Virginia Tech}
  \city{Blacksburg}
  \state{VA}
  \country{USA}
  }
\email{tauhidun@vt.edu}

\author{Zhuowei Wen}
\affiliation{%
  \institution{Virginia Tech}
  \city{Blacksburg}
  \state{VA}
  \country{USA}
  }
\email{wzhuo17@vt.edu}

\author{Brooke Ritter}
\affiliation{%
  \institution{Virginia Tech}
  \city{Blacksburg}
  \state{VA}
  \country{USA}
  }
\email{britter2@vt.edu}

\author{Shaddi Hasan}
\affiliation{%
  \institution{Virginia Tech}
  \city{Blacksburg}
  \state{VA}
  \country{USA}
  }
\email{shaddi@vt.edu}

\renewcommand{\shortauthors}{Nabi et al.}
\renewcommand{\abstractname}{\MakeUppercase{Abstract}}

\begin{abstract}
The FCC’s National Broadband Map aspires to provide an unprecedented view into broadband availability in the US.
However, this map, which also determines eligibility for public grant funding, relies on self-reported data from service providers that in turn have incentives to strategically misrepresent their coverage.
In this paper, we develop an approach for automatically identifying these low-quality service claims in the \nbmFull.
To do this, we develop a novel dataset of broadband availability consisting of 750k observations from more than 900 US ISPs, derived from a combination of regulatory data and crowdsourced speed tests.
Using this dataset, we develop a model to classify the accuracy of service provider regulatory filings and achieve AUCs over 0.98 for unseen examples.
Our approach provides an effective technique to enable policymakers, civil society, and the public to identify portions of the \nbmFull that are likely to have integrity challenges.

\end{abstract}

\ccsdesc[500]{Networks~Network measurement}
\ccsdesc[500]{Networks~Public Internet}

\keywords{Measurement, Access Networks, Policy}

\maketitle

\section{\MakeUppercase{Introduction}}
\label{sec:intro}

Access to broadband is increasingly vital in modern society for access to critical social services. 
As such, ensuring equitable access to broadband Internet is a policy priority for governments worldwide \cite{un_sdg_2015, itu_broadband_2023}.
For example, in the United States, a number of federal and state level programs exist to provide public subsidy to Internet service providers to deploy broadband infrastructure in un- and underserved areas, ranging from the Department of Agriculture's ReConnect program, the Federal Communication Commission's (FCC) Connect America Fund and Rural Digital Opportunity Fund, and the National Telecommunication and Information Administration's (NTIA) Tribal Broadband Connectivity Program (TBCP) and, most notably, the Broadband Equity, Access, and Deployment (BEAD) program, a US\$42 billion program with the goal of providing "reliable broadband" to every household in the United States~\cite{NTIA2022BEAD}.

Executing these programs requires regulators and policymakers to have an understanding of where broadband is available to enable programs to target areas with the greatest need.
As a result, alongside these programs, the FCC also maintains a map of where broadband is available~\cite{fcc_nbm}.
To do this, the FCC requires ISPs to report the locations at which they provide service every six months as part of its Broadband Data Collection (BDC) program~\cite{fccbdc}.
These reports are aggregated to produce a map which serves as the basis for determining where broadband is and isn’t available, which in turn is used by various funding programs to determine eligibility for funding.
This approach relies upon the accuracy and honesty of ISP reports, and since these are the same actors that are receiving public funds, they have incentives to misrepresent data.
A notable example is Jefferson County Cable, an Internet service provider that intentionally overclaimed the areas they covered to prevent a market just outside their territory from being eligible for public funding out of fear to minimize competition and was subsequently fined by the FCC for misrepresentation~\cite{fcc2024jefferson}.

Although the BDC program allows individuals, governments, and advocacy groups to challenge providers’ claims regarding service availability, mounting a credible challenge can be resource-intensive and costly: in the worst case, would-be challengers need to collect field data on factors such as where an operator's physical network infrastructure exists to successfully challenge provider data. 
Similarly, regulatory authorities like the FCC have responsibility for both enforcing BDC reporting requirements upon providers and providing support to providers who are striving to come into compliance with the same requirements. 
In both cases, resources are limited: the ability to both target resources for the purposes of challenges, compliance enforcement, or technical assistance, and to discriminate between the latter two, is crucial to effectively improving the quality of data collected through the BDC. 

Our goal is to develop a model to predict portions of the \nbmFull that are likely to fail if challenged; that is, claims of coverage by ISPs that are likely incorrect.
Such a model will help would-be challengers target resources more effectively and provide a means of identifying providers that have misreported their service area.
The key challenge, however, in developing such a model, is that ground truth for broadband access network availability is not readily available: mapping the physical extent and capabilities of broadband networks is difficult.
Prior work and common practice has leveraged crowdsourced speed tests~\cite{canadi2012revisiting}, measurement vantage points deployed in residences~\cite{burger2023measuring}, or provider-reported service availability~\cite{major2020no} to estimate this coverage, but each of these methods suffers from serious tradeoffs.
Crowdsourced speed tests can provide a wide breadth of independently observed coverage, but represent a self-selected population and are influenced by factors unrelated to providers' offered services.
Deploying measurement vantage points in residences is costly and difficult to scale broadly.
Reliance on broadband availability tools is still dependent on self-reported data from providers.

In this paper, we propose an alternative approach by asking the question "can crowdsourced speed tests be used to validate self-reported ISP coverage?"
To achieve this, we combine regulatory filings from the FCC's \bdcFull process with crowdsourced speed tests to produce a novel training dataset (which we will release) of broadband access network availability, consisting of more than 750,000 labelled observations from 911 service providers across all fifty states.
We demonstrate the utility of this dataset by using it to build a predictive model to identify potentially incorrect claims in the \nbmFull.
Our model performs effectively, achieving ROC AUC scores of over 0.98 for unseen samples, even holding out entire states, as well as identifying the known-false provider claim from Jefferson County Cable.

Our contributions are as follows:
    \begin{itemize}
        \item A novel dataset capturing non-archived changes to the FCC's \nbmFull
        \item A technique for associating public speed tests with providers' self-reported regulatory filings
        \item A novel dataset of broadband availability across the United States
        \item A model for predicting specific portions of the US \nbmFull that are likely incorrect
    \end{itemize}

\section{\MakeUppercase{Related Work}}
\label{s:related}
Our work builds upon a large existing literature on measuring broadband access network performance and availability.

Prior work has leveraged crowdsourced speed tests to evaluate residential broadband performance~\cite{canadi2012revisiting,deng2019estimating,paul2022importance} as well as mobile performance~\cite{mangla2022tale}.
Ookla~\cite{ookla} and MLab's NDT7~\cite{mlab} are the most commonly used crowdsourced speed tests in the literature, and we adopt these as well in this work.
These two speed tests aim to measure different aspects of network performance, with NDT7 specifically originally intended to serve as a diagnostic tool for network connections~\cite{macmillan2023comparative, clark2021measurement}.
Here, we only use data captured through these tools as an indication of the \emph{presence} of Internet availability in an area as a binary, allowing us to directly relate these datasets.
That said, we still are limited by the selection biases present in these datasets, which recent work has shown under represent locations in low-income areas~\cite{saxon2022we}.

Although we do not directly address performance in this work, a number of efforts have sought to augment regulatory broadband availability data with directly measured broadband performance from residential vantage points~\cite{staff2015ripe}. The Measuring Broadband America~\cite{burger2023measuring} program specifically aims to evaluate the longitudinal performance of residential connections at known locations on major US broadband providers.
Other efforts have similarly used vantage points at residential gateways or devices directly attached to gateways to obtain network measurements that are minimally impacted by performance bottlenecks downstream of the gateway~\cite{sundaresan2011broadband, chetty2013measuring}.
Like crowdsourced speed tests, we note that these tests do not capture what policymakers investing public funds in network infrastructure care about -- the \emph{maximum performance} available in a location, independent of what the subscriber in that location is experiencing due to choice of subscription tier or bottlenecks downstream of the provider.

Other efforts have sought to generate maps of broadband availability independently from regulatory efforts. Microsoft produced a broadband map based on data collected from product telemetry~\cite{Robinson2022, USBroadbandUsage2022} showing significantly different broadband performance compared to what was reported via the FCC's Form 477.
Most closely related to our efforts is recent work that uses providers' public broadband availability tools to estimate network coverage~\cite{major2020no} and pricing~\cite{paul2023decoding}.
While this work relies upon self-reported data from providers, these consumer-facing tools similarly appear to report different data than providers report as part of their regulatory filings.
Such discrepancies motivate our work, and we also find these alternative approaches complementary to our own: observations from the techniques proposed in these related efforts could be used to enhance our broadband availability dataset.

\section{\MakeUppercase{Building the US \nbmFull}}
\label{s:background}

We begin by providing background on the FCC's \bdcFull program, which provides the data that underlies the current US \nbmFull.
We focus here on mapping \emph{fixed} broadband service; the FCC maintains a separate though analogous process for mapping mobile coverage that we do not consider in this work.

\textbf{Legislative and Regulatory Background.}
In the wake of the Telecommunications Act of 1996~\cite{telecommunications1996}, the FCC began its first program for collecting data about the state of broadband deployment in the United States, establishing the Form 477 data collection program in 2000~\cite{fcc2000report}.
Form 477 required ISPs to report the postal codes in which they provided broadband service on a twice-yearly basis; the Commission increased the reporting granularity to the census tract level in 2008~\cite{fcc2008report} and further to the census block (the smallest statistical unit used in the United States) in 2013~\cite{fcc2013form477}.
Form 477 was initially motivated by the need to understand broadband availability and competition in the US, but would evolve to form the basis of the first US National Broadband Map developed by the NTIA in 2011.

Data from Form 477 faced criticism due to the coarse and varying spatial resolution of census blocks, leading to overestimates of service availability, particularly in rural areas~\cite{rochester2018broadband}.
As a result, in 2020 the US Congress passed the Broadband DATA Act~\cite{broadbanddataact2020} to direct the the FCC to develop a new broadband data collection program that would collect data at the granularity of \emph{Broadband Serviceable Locations} -- effectively individual buildings -- and to create a process to solicit public feedback on data reported by ISPs -- the \emph{challenge process}.
This new process, known as the \bdcFull, began collecting data in 2022 and resulted in the first publication of the current \nbmFull in November 2022.
Beyond providing a more granular view into broadband availability, the new \nbm would also serve as the basis for determining eligibility for funding from the NTIA's BEAD program~\cite{NTIA2022BEAD}, with each US state receiving a share of the program's US\$42B budget in proportion to the number of locations lacking reliable broadband.

\textbf{Broadband Serviceable Location Fabric.}
The current \bdc process starts from a semi-public dataset known as the \emph{Broadband Serviceable Location Fabric}, often simply referred to as the Fabric.
The Fabric is intended to capture every structure in the United States that could be served by broadband (regardless of whether or not it currently is); each structure in the fabric is a Broadband Serviceable Location (BSL).
Each BSL is a point with metadata associated with the structure it represents, such as address, GPS coordinates, number of units (for multi-dwelling buildings like apartments), as well as designations for BSLs that are "community anchor institutions" such as schools, libraries, and hospitals which are classified separately as part of the \bdc process.\footnote{The FCC assumes that such locations do not subscribe to "mass market" consumer or business broadband, and further are eligible for different sets of public funding programs.}

Critically, ISPs can only report service to locations present in the Fabric: if a structure is not represented in the Fabric, no operator can claim service there, even if they already provide service at that location.
The public may suggest changes to the Fabric to add missing locations or correct information already present in the Fabric; these changes are reviewed by the FCC and CostQuest, the vendor who develops the BSL Fabric, every six months to produce a revised version of the BSL Fabric used in the \bdc process.

\textbf{Generating the \nbmFull.}
To generate the \nbmFull, the FCC requires all ISPs in the US to submit data about the BSLs they serve every six months as part of the \bdc process.
Each reporting period uses a new version of the BSL Fabric.
ISPs provide a list of BSLs they serve or could serve if requested within ten business days, and, for each BSL, the maximum speed they offer and the type of technology they use to serve the location.\footnote{The FCC also allows ISPs who provide service using wireless technologies to provide a polygon representing their coverage areas, and will calculate the locations within that polygon on the ISPs behalf. However, in practice, very few ISPs do this as this option requires ISPs to submit extensive technical documentation about their network configuration, making this more burdensome than simply providing a list of locations.}

Table~\ref{t:bdc_data} summarizes this data.
Providers also submit a free-text description of the methodology used to determine which locations were served versus unserved; this methodology is public as well.
The FCC gives discretion to ISPs to determine what they consider "served": as we will discuss, ISPs use a range of methodologies to determine their service area, ranging from propagation models, engineering information about fiber routes, or simply subscriber addresses; similarly, ISPs have discretion as to what service speeds they advertise at each location.
In addition, ISPs provide counts of active subscribers per technology and speed plan offered per census block, similar to the Form 477 data previously required.

Finally, to generate the \nbmFull, the FCC aggregates this data collected from ISPs, typically publishing a new version of the \nbm 4-5 months after the ISPs' filing deadline.
The public version of the data that comprises the NBM is simply a list of each ISP's claims of service for each BSL, as well as the census block and an H3~\cite{h3geodocs2023} grid identifier in which that BSL is located.

Since its initial release in 2022, the \nbm has undergone regular major updates associated with each new ISP filing deadline and corresponding Fabric version (roughly every six months) as well as minor updates roughly every two weeks capturing changes made to the initial release of the map since the last major version.

\begin{table*}[ht]
    \centering
    \begin{tabular}{>{\raggedright\arraybackslash}p{0.29\textwidth} 
                    >{\centering\arraybackslash}p{0.08\textwidth} 
                    >{\raggedright\arraybackslash}p{0.6\textwidth}}
         \toprule
         \textbf{Item} & \textbf{Unit} & \textbf{Notes} \\
         \midrule
         Max Advertised Download Speed & Mbps & Values below 10Mbps are reported in the NBM as 0. \\
         Max Advertised Upload Speed & Mbps  & Values below 1Mbps are reported in the NBM as 0. \\
         Latency $\leq$ 100ms & Boolean & Latency measurement methodology is not defined. \\
         Access Technology & Category & Copper, fiber, cable, GSO/NGSO satellite, and licensed/unlicensed wireless. \\
         Service Type  & Category & Business, Residential, or Both. \\
         \bottomrule
    \end{tabular}
    \vspace{4pt}
    \caption{Summary of data ISPs submit to the FCC every six months as part of the \textit{\bdcFull} process for each location they serve with fixed broadband service. Providers also report a "brand name" for the service, which is typically the name of the ISP, but can vary, e.g., Comcast reports under their "Xfinity" brand.}
    \label{t:bdc_data}
    \vspace{-.2in}
\end{table*}

\textbf{Correcting the \nbmFull.}
Once the \nbm is published, individuals and organizations can dispute (or "challenge") availability claims of ISPs; this drives many of the changes seen in the "minor" updates to the \nbm published every two weeks.
At a high level, once a challenge is filed with the FCC, after an initial preliminary review by FCC staff, the challenge is forwarded to the relevant ISP.
The ISP has 60 days to review the challenge to decide whether to concede or dispute the challenge; if the provider disputes the challenge, both the challenger and the provider have an additional 60 days to correspond to reach an agreement.
If after this second 60 day period the parties still disagree, the FCC then takes responsibility for evaluating evidence provided by both parties to adjudicate the challenge within the next 90 days; the FCC's determination is final.
Thus, a challenge may be resolved and updated on the map in as little as two weeks (e.g., if the provider immediately concedes the challenge) or after more than seven months for challenges that require FCC adjudication.

The burden of evidence required for a challenge varies based on the scope of the challenge.
The FCC is required by the Broadband DATA Act to develop a "consumer friendly" challenge process~\cite{broadbanddataact2020}; thus, an individual challenging a single location via the FCC's online system need only provide a high-level description of why they are disputing a provider's availability claim to initiate a challenge.
In contrast, organizations such as state broadband offices, Tribal or local governments, and researchers seeking to challenge a provider's claim at multiple locations (a "bulk challenge") must provide specific evidence in support of their challenge at \emph{each location} to pass the initial FCC review and initiate the challenge process.
Importantly, \emph{the FCC does not consider broadband speed tests sufficient evidence to initiate a challenge}, instead requiring knowledge of physical infrastructure or data other than speed tests collected from individual consumers.
This requirement makes bulk challenges costly, which motivates our work.

\section{\MakeUppercase{Data}}
\label{s:datasets}
Given this background, we now turn to using the data collected as part of the \bdc process to develop a dataset approximating "ground truth" for broadband availability in the US.
Our goal is to develop a dataset of locations for which we have high confidence regarding whether they are served or unserved by a particular network operator.
Our insight is that such a dataset would comprise a set of labels for training models of broadband availability for downstream tasks such as evaluating the quality of ISP coverage claims; we develop such a model in Section~\ref{s:model}.

Our approach proceeds as follows.
First, we derive approximate geographic locations of every BSL using only publicly available data.
Next, we leverage a subset of data collected in the \bdc process that is \emph{not} solely reliant on self-reported data from service providers: service claims that have been challenged by the public or through the FCC's internal data quality checks.
We then augment these claims with crowdsourced speed test data to derive a synthetic set of locations that are very likely to be actually served by a provider (Section~\ref{s:datasets:synthetic}).
Finally, we use these inputs to develop a labelled dataset of broadband availability to be used for further downstream tasks.
We focus on the initial release of the \nbmFull first published in November 2022, reflecting broadband deployment as of June 30, 2022 (the corresponding ISP filing deadline).

\subsection{\bdcFull}
\subsubsection{Localizing BSLs}
\label{s:datasets:location}
We begin by localizing BSLs. Under CostQuest's license terms for their Fabric dataset, we cannot use the precise GPS coordinates of BSLs; researchers are restricted to reporting results that use the Fabric aggregated to the county level.
As a result, we cross reference the H3 grid (resolution level 8)~\cite{h3geodocs2023} location of each BSL that appears in the \nbm, which is fully public; these grid cells are approximately $0.5 km^2$ in area.
Although this approach does not allow us to identify the location of BSLs that are unserved by any provider and thus do not appear in the \nbm, we found that, on a county level, only seven counties nationwide have a different number of BSLs in the \nbm compared to the underlying Fabric dataset.
These seven counties are all either part of Pacific Ocean territories or rural municipalities of Alaska.
This correspondence comes from the fact that geostationary satellite providers report service to essentially every location in the United States, and these records appear in the \nbm.
This gives us confidence that we are capturing the H3 grid location of the vast majority of BSLs available publicly.
However, this also creates limitations for our results, as it limits our unit of analysis in this work to these H3 grid cells at resolution level 8, rather than individual locations.

\subsubsection{\bdc Challenges}
\label{s:datasets:challenges}

The FCC publishes the outcomes of challenges to the \nbm on a monthly basis, noting one of five primary outcomes (Table~\ref{t:challenge_outcomes}). Successful challenges are those that succeed in removing or modifying data reported by an ISP from the \nbm; failed challenges represent cases where an ISP is able to defend its claims. Challenges are resolved either through mutual agreement between the challenger and ISP or, in the absence of agreement, by the FCC through a review of submitted evidence.
Challenges are useful as they represent a subset of provider claims that are not dependent solely on self-reported data alone; at least one other party (the challenger or the FCC) has evaluated the claim at those locations.
Indeed, in most cases successful challenges are due to a provider conceding that their initial claim was incorrect.
Conversely, failed challenges represent cases where the challenger abandoned their challenge after discussion with the provider or, more commonly, the FCC found insufficient evidence to rebut a claim.
We treat successful challenges as an indication that providers do not provide service in a hex, and failed challenges as evidence that the provider's claim is valid.

\begin{figure*}[h!]
  \centering
  \includegraphics[width=\textwidth]{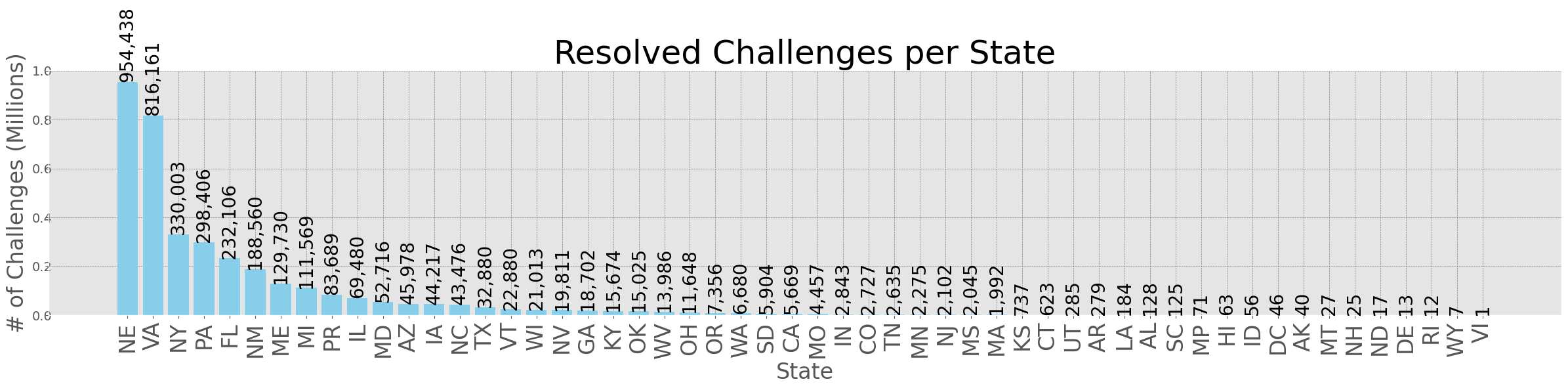}
  \caption{The state-by-state challenges to the initial Broadband Data Collection (BDC) filings as of June 30, 2022, show significant disparities. A few states contribute nearly 1 million challenges, while many other states report fewer than a thousand challenges.}
  \label{fig:state_wise_challenge}
\end{figure*}

Unfortunately, challenges alone are insufficient to create a training set for our model as they do not constitute a representative sample of provider service claims.
Public challenges in the time period we evaluate are heavily weighted to specific states (Figure~\ref{fig:state_wise_challenge}), with just ten states accounting for around 90\% of challenges.
This is a consequence of the policy incentives of challengers: the NTIA's BEAD program was designed to allocate funding to state broadband offices proportionately to how many un- and underserved locations were in each state, based on data in the \nbm.
As a result, some state broadband offices engaged heavily in the challenge process as this would directly increase the amount of funding the state would receive from this program (one such example is Virginia, where an extensive challenge campaign increased the state's BEAD allocation by more than US\$250 million~\cite{fleming_2023}).
Further, given the effort required for individuals or organizations to file challenges, it is reasonable to expect that the set of locations challenged is biased towards locations where challenges are likely to be successful; indeed, almost 70\% of challenges succeed (Table~\ref{t:challenge_outcomes}).

\begin{table}[]
    \centering
    \begin{tabular}{@{}p{0.22\textwidth}@{\extracolsep{4pt}}rr@{}}
    \toprule
    \textbf{Challenge Outcome} & \multicolumn{1}{c}{\textbf{\# BSLs (\%)}} \\
    \midrule
    \textbf{Successful} & \textbf{2,499,012 (69\%)} \\
    \quad Provider Conceded & 1,408,171 (39\%) \\
    \quad Service Changed & 814,432 (22\%) \\
    \quad FCC Upheld & 276,409 (8\%) \\
    \midrule
    \textbf{Failed} & \textbf{1,124,470 (31\%)} \\
    \quad Challenge Withdrawn & 533,494 (15\%) \\
    \quad FCC Overturned & 590,976 (16\%) \\
    \bottomrule
    \end{tabular}
    \vspace{4 pt}
    \caption{Distribution of challenge outcomes for the initial release of the \nbm. Challenges began in November 2022 and continued through November 2023 on the initial NBM. Most challenges succeed, indicating the provider's original service claim was incorrect.
    }
    \label{t:challenge_outcomes}
\end{table}

\begin{table}[h!]
    \centering
    \begin{tabular}{@{}p{0.36\textwidth}@{\extracolsep{\fill}}r@{}}
    \toprule
    \textbf{Reason for Challenges} & \multicolumn{1}{c@{}}{\textbf{Counts (\%)}} \\
    \midrule
    Technology Unavailable & 1,984,512 (55\%) \\
    Speed(s) Unavailable & 1,555,178 (43\%) \\
    Service Request Denied & 46,048 (1\%) \\
    No Signal & 36,945 (1\%) \\
    Asked Higher than Standard Connection Fee & 361 (<1\%) \\
    Failed to Provide Service within 10 Biz-days & 254 (<1\%) \\
    Provider not Ready (dependency on new equipment) & 99 (<1\%) \\
    Failed to Install Service within Timeline & 85 (<1\%) \\
    \bottomrule
    \end{tabular}
    \vspace{4pt}
    \caption{Distribution of reasons for challenges in the initial release of \nbm. Nearly all challenges were due to the reported network infrastructure being unavailable at a location ("Technology Unavailable") or a provider not offering claimed service speeds ("Speed(s) Unavailable").} %
    \label{t:reasons_for_challenges}
\end{table}

\begin{figure}[h!]
  \centering
  \includegraphics[width=\linewidth]{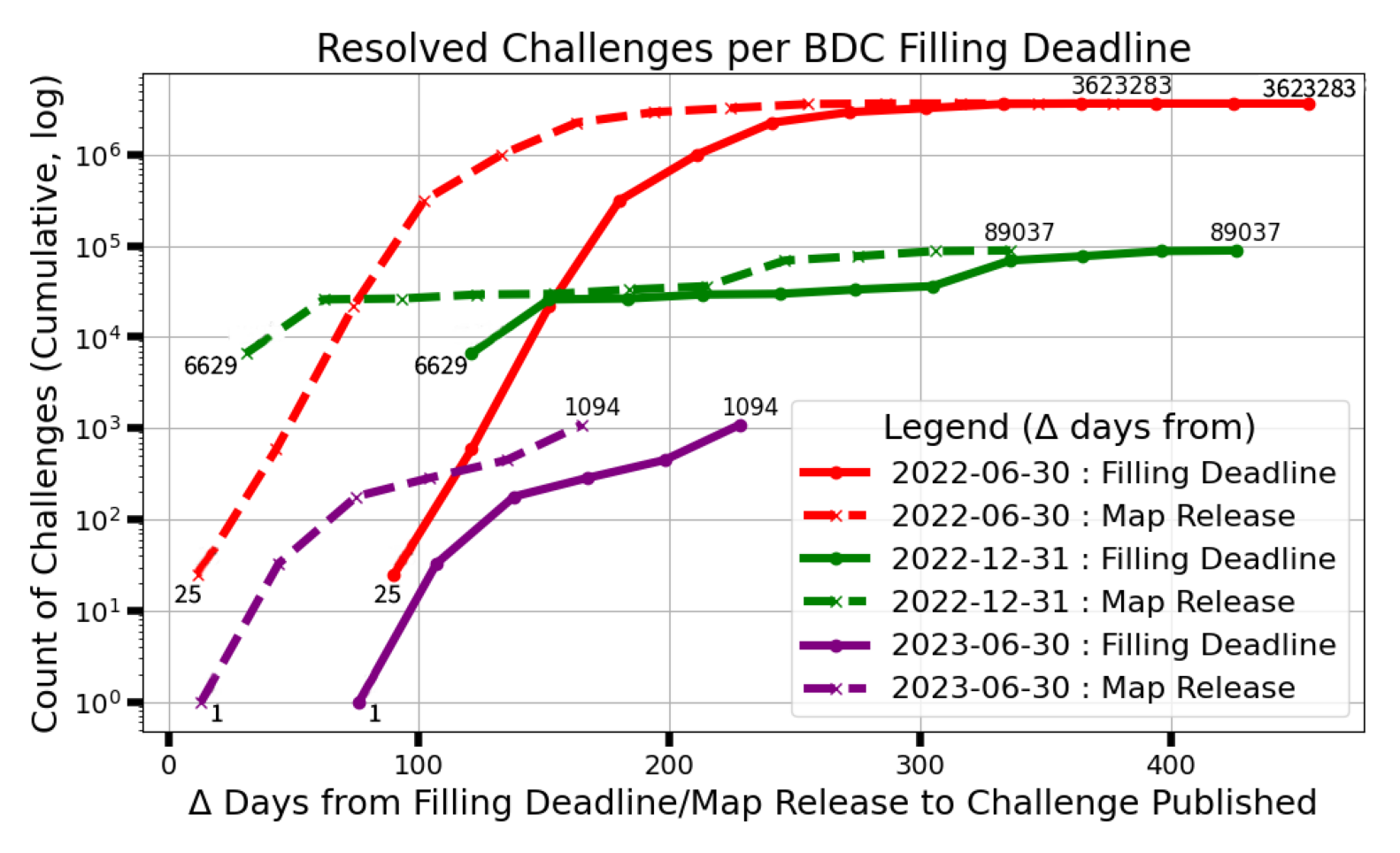}
  \caption{Count of challenges over time for each major and minor release of the \nbm. The first major release, which we focus on in this work, saw nearly two orders of magnitude more challenges than the subsequent release.}
  \label{fig:submission_wise_challenge}
\end{figure}

\subsubsection{Augmenting Public Challenges with Non - Archived Changes}
\label{s:datasets:change}
The public challenge process is not the only process by which provider claims are reviewed.
Although only the locations specifically challenged are released by the FCC, in responding to a challenge a provider may choose to modify other aspects of their filing -- e.g., if a challenge reveals a methodological error in the original filing.
Similarly, the FCC also conducts its own internal quality checks for data submitted by providers~\cite{fcc_broadband_data_task_force_2023}, proactively engaging with operators to update their filings.
Both of these processes would result in \emph{changes} to the data reported in the minor bi-weekly updates to the \nbm and provide another view into inaccuracies of initial provider service claims.

To capture these changes, we captured releases of the \nbmFull starting in December 2022\footnote{Our initial capture occurred four weeks after the \nbm's initial public release. We began archiving all major and minor map updates in Feb 2023 once we discovered the \nbm was being updated every two weeks between major updates.} and computed the changes in each provider's claims between their initial filings and the most recent minor release of the map.
These "map diffs" allow us to observe changes to provider filings that are not directly attributable to formal challenges and include providers both adding or removing locations from their service claims, as well as modifying reported service offerings in locations.

For the purposes of our building our dataset for labeling broadband availability we only consider locations \emph{removed} from a provider's initial claim.
Any locations removed we treat as successful challenges, under the assumption that such removals were prompted by FCC-initiated quality checks or in response to a public challenge revealing a methodological flaw in the initial filing.

By incorporating these inferred challenges, we identify 185k additional observations where ISPs provided incorrect information, beyond those in public challenges.

\subsection{Inferring Likely Served Locations}
\label{s:datasets:synthetic}
The validated data we can extract from the \bdc process largely provides a view of locations that were incorrectly reported in providers' \bdc filings: the \bdc process provides mechanisms for allowing third parties to dispute a providers' filing, but none for validating that ISPs are accurately reporting the service provided in a location.

To address this, we turn to crowdsourced speed tests from Ookla~\cite{ookla} and MLab's NDT7~\cite{mlab}.
Each of these datasets provides longitudinal data on network performance collected from client devices globally, collectively comprising billions of individual active measurements.
While these tests can provide a useful view into access network performance, they have well-known limitations for assessing whether an ISP provides service at a location~\cite{paul2022importance}, including the fact that such tests can indicate performance bottlenecks due to subscription tier, in-home network performance, or even the speed test methodology itself~\cite{feamster2020measuring}.
More fundamentally, these speed test datasets, much like \nbm challenges, represent a convenience sample of locations from which a self-selected set of users chose to run the tests~\cite{clark2021measurement}.

As a result, we do not directly compare speed test results to provider claims in the \nbm.
Instead, we only use speed tests to identify locations where, with high probability, a provider actually provides service as claimed.
We make only a weak assumption that \emph{the existence of a speed test in a location suggests that Internet service is available there}; we refer to the set of locations that we identify this way as \emph{likely served locations}.
A consequence of this is that, because we only consider the existence of speed tests, we can combine results from datasets that otherwise use incomparable methodologies.

\subsubsection{Crowdsourced Speed Test Datasets}
\label{s:datasets:synthetic:cds}

We first describe the underlying speed test datasets we use, Ookla's Open Data Initiative~\cite{ookla_data} and MLab's NDT7~\cite{mlab}.

\noindent\textbf{Ookla Open Data Initiative.} The Ookla Speedtest~\cite{ookla} is an active measurement conducted between a users on their web or mobile client accessing their network of more than 15,000 speed test servers, resulting in more than 11 million daily tests~\cite{ookla_network}.
These tests capture a number of network performance parameters including upload and download throughput as well as latency.
Ookla releases a public, aggregated version of this data each quarter~\cite{ookla_data} consisting of only those speed tests where the client provides precise GPS location. For example, a mobile phone on a WiFi network running the Ookla native Speedtest app with location permissions enabled would be included, whereas a test run from a browser client without location data would not.
The public version of this dataset is aggregated into "quadkey" tiles~\cite{bing_maps_tile_system}, each representing a rectangular grid approximately 500m on a side.
For each such tile, Ookla provides the count of unique tests, count of unique devices, mean download and upload throughputs, and mean latency, aggregated across all providers.
We re-project this Ookla dataset from quadkeys to H3 grid tiles to align with our other datasets (Appendix~\ref{s:appendix:hex_system}).

\noindent\textbf{MLab NDT7.}
MLab's NDT7~\cite{mlab} test is also an active measurement conducted between clients, typically via their web interface, and a network of servers operated by MLab.
Unlike Ookla, all NDT7 tests are available publicly; beyond throughput and latency, each test also includes the client's source IP address and ASN.
MLab does not capture user location, but uses IP geolocation to provide an estimated location and accuracy radius for each test~\cite{gill2022mlab}.

\subsubsection{Attributing and Localizing MLab Speed Tests}
\label{s:datasets:synthetic:loc_mlab}

We next identify the set of hexes from which an MLab test may have been conducted.
To do this, we first \emph{attribute} MLab speed tests to individual providers that completed \bdc filings, and then we \emph{localize} each speed test to the intersection of hexes claimed by the provider and within the accuracy radius provided by MLab's IP geolocation.

\noindent\textbf{Provider attribution.} 
Each provider that participates in the \bdc process has a unique Provider ID that is in turn associated with one or more FCC Registration Numbers (FRNs)~\cite{fcc_bdc_provider_id_table}.
FRN registration data includes information on the legal entity associated with each FRN, including contact information and location.
We match this information against ARIN's WHOIS registration data~\cite{arin_bulk_whois} to identify match ASNs to Provider IDs (Appendix~\ref{s:appendix:asn_mapping_methodology}).
 
We evaluate our matching procedure in greater detail in Section~\ref{s:eval:provider_asn}, but we find strong correspondence between the provider and ASN mappings we obtain through our four independent matching techniques, and are able to match 72\% of providers that completed filings to ASNs.

\noindent\textbf{Test localization.}
After mapping providers to ASNs, we use the ASN associated with each MLab speed test to localize it within the claimed service footprint of the corresponding provider.
To do this, we first identify a set of H3 hexes in which the MLab test \emph{may} have been conducted; these are all hexes within the accuracy radius recorded in the IP geolocation of the test.
We exclude all tests with accuracy radius of more than 20km to reduce the candidate set of locations.
Then, we intersect this candidate set of hexes with the hexes that the corresponding provider claimed in the \nbm.

From this, we produce a set of H3 grid cells from which MLab speed tests could have been run for each provider for which we are able to attribute ASNs. Although MLab exposes significantly richer data than Ookla's public dataset, we leave further exploration to future work.

\subsubsection{Generating Synthetic "Likely Served Locations"}
\label{s:datasets:synthetic:known_goods}

Using these crowdsourced speed tests, we then proceed to identify areas that are both claimed by a provider in the \nbm and likely to be served by that provider.
Our intuition is that \emph{areas with high density of speed tests that can be attributed to a particular provider are more likely to be actually served by that provider.}
Our goal is to identify these locations to use synthetic "known good" service claims from providers during training.

To do this, we first use Ookla data to identify all H3 cells with a ratio of unique devices per BSL greater than one; we refer to this ratio as a service coverage score for a H3 cell.
Intuitively, these hexes had at least one unique Ookla device conduct a speed test per BSL, suggesting that service is widely available within the cell from \emph{some provider}; Ookla data alone does not allow us to identify specific providers.
We then take these cells and intersect the set with the set of cells identified from attributed and localized MLab tests for each provider (as described in Section~\ref{s:datasets:synthetic:loc_mlab}).
These MLab derived maps allow us to associate providers with each candidate hex, and we are further able to associate technology types with each candidate hex from provider claims in the \nbm.

Using this methodology, we can generate candidate synthetic "known good" provider claims where
(a) 
Ookla data indicates a high volume of speed tests (from some provider) at that location,
(b) 
M-Lab data suggests a speed test may have been conducted in the hex from a specific provider’s network, 
and 
(c) 
that provider claims in the \nbm to offer service in a location using a given technology.

\subsection{Defining Labelled Observations}
\label{s:datasets:labels}
Finally, we produce a labelled dataset of observations upon which we can train our model.
We tackle this challenge by initially using data from challenges (Section~\ref{s:datasets:challenges}) and progressively augmenting it with non-archived "change" data (Section~\ref{s:datasets:change}) and our likely served locations. (Section~\ref{s:datasets:synthetic}). 

We define an observation to be a combination of provider, H3 resolution 8 hex cells, and technology type.
This definition arises naturally from the \nbm: each provider may report service with multiple technologies per location (effectively producing one filing per technology type), and providers' claims are reported publicly at the H3 resolution 8 level.
We define binary labels for each observation to reflect whether that observation is \emph{served} or \emph{unserved}; note by this definition, a single hex can be labelled both served by some providers and unserved by others.

To assign labels, we proceed as follows.
First, any observations which were successfully challenged are labelled as unserved, while any observations that were \emph{unsuccessfully} challenged are labelled as served.
Next, any locations that were removed from an original filing as identified from changes in the public \nbm are labelled as unserved as well.

We then attempt to balance the resulting dataset on a per-provider, per-state basis using our set of inferred likely served locations ordered by descending service coverage score we defined in Section~\ref{s:datasets:synthetic:known_goods}.
We attempt to balance observations on a per-state basis because we know the process that generated challenges occurred on a per-state basis.
Similarly, each provider chooses its own methodology and strategy for defining its service claims via the \bdc process as well as its own policy for responding to challenges.
When it is not possible to generate sufficient likely served location observations to balance each provider, we attempt to balance the state as a whole.

We emphasize balancing our dataset because a wide range of machine learning literature acknowledges that an imbalanced dataset can significantly hurt model performance\cite{cateni2014method} or lead to unfair and inaccurate predictive models~\cite{zhu2022consistent}.
In our study, we observed similar issues when running our model on a highly imbalanced dataset. The performance suffered due to the skewed class distribution, which led to poor generalization and misclassification of the minority class. 

Our analysis time frame focuses on the initial \bdc filing, reflecting network deployments as of July 30, 2022, for which the corresponding \nbmFull was released in November 2022.
We chose this initial release given the significant number of challenges and changes available compared to later releases (Figure~\ref{fig:submission_wise_challenge}).
Thus, for both the Ookla and MLab datasets, we consider data from October 2021 through September 2022.
For the "change" data, we focus on the deletion and modification recorded for the July 30, 2022 map during the period from February 2023 through November 2023.\footnote{Due to a data collection error, our Feb 2023 snapshot of the \nbm is our first complete map; our December 2022 snapshot omitted New York.}
We similarly include challenges beginning in February 2023 and discard all previous challenges from our training to avoid data leakage.

Our final labelled dataset consists of approximately 750,000 observations covering filings from 911 operators,\footnote{This is less than the number of providers with matching ASNs, as only a subset of matched ASNs had usable MLab test results.} covering 545,000 hex cells in all 50 states.  The labelled dataset is composed of observations derived from challenges (51\%), non-archived changes (22\%), and synthetic likely served locations (27\%).\footnote{This data is available at \url{https://github.com/spin-vt/red_is_sus}.}

\section{\MakeUppercase{Modeling \nbm Integrity}}
\label{s:model}

We now turn to using our labelled dataset of broadband availability to train a classifier for evaluating the integrity of \bdc filings.
Specifically, we are interested in classifying which portions of a \bdc filing by a provider are likely to fail challenges; that is, of the locations a provider claims to serve, which of these are most likely to be unserved.

\subsection{Feature Engineering the \nbm}
\label{s:model:features}

As described in Section~\ref{s:datasets:labels}, our unit of observation is $(\text{H3 cell},\allowbreak \text{access technology}, \text{provider})$.
We augment this vector with additional data obtained from the \nbm as well as datasets described in Section~\ref{s:datasets}, as described in Table~\ref{t:vectorization}.
Specifically, for each observation, we include the maximum advertised upload and download speeds, state/territory, centroid of the H3 cell, and percentage of locations for which the provider claims service in the cell.
If a provider claims multiple speeds in a cell, we pick the maximum advertised download speed and corresponding upload speed.
In addition, we include an embedding~\cite{sbert} of the free-text methodology provided by the ISP in their \bdc filing.

We also incorporated crowd-sourced test attributes (section \ref{s:datasets:synthetic}) into the model. Instead of using speed results directly, we considered Ookla and MLab attributes such as test counts and device data per H3 location. We avoided direct incorporation of third-party speed test results as they may reflect in-home conditions, which are not comparable to the maximum advertised speeds by ISPs. Other test attributes provide a more reliable measure of internet availability in these areas.

\begin{table}[h!]
    \centering
    \begin{tabularx}{\columnwidth}{>{\raggedright\arraybackslash}X >{\raggedright\arraybackslash}X >{\raggedright\arraybackslash}X}
    \toprule
    \textbf{Features} & \textbf{Vectorization} & \textbf{Notes} \\
    \midrule
    Maximum advertised speed (downlink and uplink) & Maximum reported value & Max of max advertised speed \\
    \hdashline
    Low latency & Binary representation & Indicates presence or absence of low latency \\
    \hdashline
    State & One-hot encoded \cite{breskuviene2023categorical} state tag & Encodes state information \\
    \hdashline
    Location & Centroid GPS coordinates for each hexagon & Provides geographical positioning \\
    \hdashline
    Location Claims & Percentage of location claims & Overview on ISP's coverage footprint \\
    \hdashline
    Methodology & S-BERT~\cite{sbert} Embedding & Overview of the methodology followed by ISP \\
    \hdashline
    Ookla Tests & Device(s) per locations in a H3 & Availability of third party speed tests \\
    \hdashline
    MLab Tests & Test counts per ISP in a H3 & Availability of third party speed tests \\
    \bottomrule
    \end{tabularx}
    \caption{Overview of how the vectorization technique is applied to \bdc filing attributes}
    \label{t:vectorization}
    \vspace{-0.2in}
\end{table}

This vectorization approach allows us to generate approximately 13 million observations across the United States, based on the initial BDC submissions from 2,153 providers, excluding claims from 3 non-terrestrial (satellite) service providers.
Because we are only able to generate synthetic likely served locations for providers we are able to find ASN matches and MLab speed test for, we consider only the subset of 911 providers.

\noindent\textbf{Incorporating Provider Filing Methodology.}
As part of the \bdc process, each provider submits a public, free-text methodology describing how they generate their service availability reports for each technology \cite{fcc2023bdcspec}.
Manually inspecting these methodologies revealed wide variation in approach.
Some providers reported methodologies explicitly disallowed by the FCC, such as simply reporting full census blocks of coverage as done for the earlier Form 477 data collection effort.
Similarly, we noticed sets of ISPs that all used nearly word-for-word identical methodologies, which appeared to be due to the use of the same consultants preparing filings on behalf of many ISP clients. Building on this observation, we use S-BERT~\cite{sbert} to generate vector embeddings for each providers' methodology statements, which we append to every observation for a provider.
We chose vanilla S-BERT for this task due to its robust performance in generating contextual embeddings, which capture semantic similarities between sentences effectively. The pre-trained S-BERT model allows us to generate 384-dimensional vectors for each observation, which we also used as a feature to train our model.

\subsection{Model Architecture}
\label{s:model:xgboost}

Interpretability was a priority in our choice of model selection given the regulatory relevance of our work.
As such, we use a tree-based algorithm for developing our classifier due to its easy interpretability, ability to effectively capture non-linear relationships, and flexibility in handling various data types, such as numerical and categorical \cite{breiman2001random, jost2006tree}.
Specifically, we use the eXtreme Gradient Boosting (XGBoost) algorithm, known for its sparsity awareness and enhanced regularization to prevent overfitting—a significant advantage over other tree-based algorithms \cite{xgboost}; the algorithm scales well and enables fast model development~\cite{ramraj2016experimenting}. 

While building our XGBoost model, we used standard hyperparameters, including learning rate and cross-validation, to ensure regularization and avoid over-fitting~\cite{santos2018cross}. We optimized hyperparameters using Bayesian optimization to efficiently explore the hyperparameter space~\cite{pelikan2005bayesian}.

In Appendix \ref{s:appendix:interpretation}, we focus on the interpretations of the XGBoost model results by analyzing the factors influencing the model's predictions and characterizing its outcomes.

\section{\MakeUppercase{Evaluation}}
\label{s:eval}

\subsection{Provider to ASN Mapping}
\label{s:eval:provider_asn}
Because of the centrality of MLab's speed tests to our approach, both as a means of identifying likely served locations as well as in our model, we begin by evaluating our methodology for mapping providers to ASNs.

\begin{figure}[h!]
  \centering
  \includegraphics[width=0.85\linewidth]{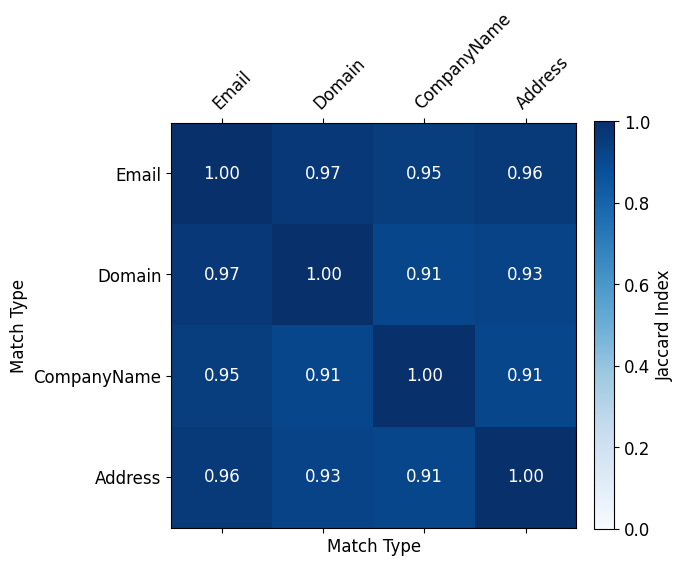}
  \caption{Mean Jaccard Index matrix for provider to ASN mappings by methodology.}
  \label{fig:asn_mapping_jaccard_similarty_matrix}
\end{figure}

To assess the accuracy of our ASN to provider mappings, we consider agreement among our four independent techniques for comparing ASN registration data to FCC registration data: full email address, physical address, company names, and contact email domain.
High agreement across these distinct methods increases confidence that our mappings are both correct and complete.
We observe that providers often have a \emph{set} of associated ASNs -- for example, Comcast's primary ASN is AS7922, but they also are associated with 58 other ASNs. 
Figure~\ref{fig:asn_mapping_jaccard_similarty_matrix} depicts the Jaccard Index~\cite{jaccard1901} for provider to ASN mappings across each of our matching methodologies.
In general, we see a high degree of agreement across our mapping methods, with most of our matches coming from email domain or company name.

In total, we are able to identify ASNs for 1562 of 2156 (72\%) of providers.
Of these, 944 are strong matches, with matches obtained from multiple methods and a Jaccard Index of 1.
An additional 171 providers have matches across multiple methods with a Jaccard Index of <1, and the remaining 447 obtained matches from only a single method.

Interestingly, we also found 226 ASNs that mapped to multiple providers.
We randomly sample 20\% of these to manually inspect these providers to determine the quality of our matches, comparing registration metadata and publicly available information about corporate entities to understand the relationships among providers that share ASNs.
The majority of these shared ASNs are either linked to various entities that appear to be part of the same corporate group or rely on the same regional wholesale IP transit provider.

\begin{table}[]
    \centering
    \begin{tabular}{lr}
    \toprule
    \textbf{Matching Methodology} & \textbf{\# Providers} \\
    \midrule
    Full Email Address & 293 \\
    Contact Email Domain & 1173 \\
    Company Name & 1163 \\
    Physical Address & 729 \\
    \bottomrule
    \end{tabular}
    \caption{Overview of providers matched to ASNs by method. Of a total of 2156 providers in the \nbm (2022-06-30), we found at least one corresponding ASN for 1562 (72.4\%).}
    \label{t:provider_id_matching}
    \vspace{-0.2 in}
\end{table}

\begin{figure}[h!]
  \centering
  \includegraphics[width=0.85\linewidth]{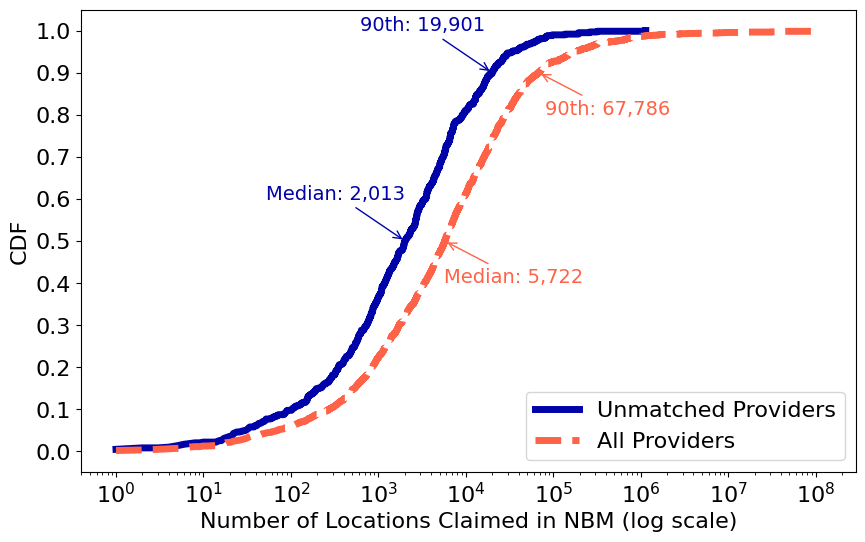}
  \caption{CDF of locations claimed in the {\nbm} by unmatched providers and all providers. The median and 90th percentile of location claims for all providers are approximately three times higher than those of unmatched providers.}
  \label{fig:cdf_unmatched_providers_location}
\end{figure}

Finally, we evaluate the set of 583 providers for which we were \emph{not} able to identify any matching ASNs.
We do not expect every provider to have an ASN -- for example, small providers may be single-homed and not have an ASN of their own. 
Indeed, we find that most of our unmatched providers are small: the median unmatched provider claims 2013 locations, and the 90th percentile claims 19901 locations (Figure~\ref{fig:cdf_unmatched_providers_location}); in comparison, the median and 90th percentile ISP in the overall \nbm are approximately three times larger.

Although we hope to reduce this set in future work, our methodology allows us to associate most providers in the \nbm with ASNs, which in turn allows us to associate public network measurements with their filings.

We also evaluate our provider attribution approach against existing AS sibling relationships datasets such as as2org~\cite{as2org} and as2org+~\cite{as2orgplus}, since they provide valuable insights in network topology. Even though our approach isn’t intended to capture AS sibling relationships, in practice it achieves a similar result for the subset of ISPs reporting data to the \nbm. We compared the ASNs groupings we identify with the groupings from as2org+ and find a high degree of overlap: we observe a mean Jaccard Index~\cite{jaccard1901} of $\approx 0.9$ between corresponding ASN groupings, with 1243/1562 ASN groupings we identified exactly matching an ASN grouping from as2org+.

\subsection{Model Performance}
\label{s:eval:performance}

We now evaluate our model's performance predicting the likelihood of challenge failures based on the training data we developed in Section~\ref{s:datasets}.
Our approach focuses on evaluating the utility of our model in providing \emph{useful} predictions for would-be challengers of the \nbmFull.
Because we do not have access to ground truth, we employ an evaluation strategy that compares our models' performance to a naive "random guessing" approach as a baseline for predicting challenge outcomes.
We evaluate performance by dividing our training dataset into three distinct subsets, training, testing, and validation. This last subset remains completely unseen during both training and validation phases and is used solely for performance evaluation.
We choose our holdout sets to correspond to scenarios organizations preparing challenges to the \nbm may face; the training and testing datasets correspond to 90\% and 10\% of the remaining non-held-out labeled data.

\subsubsection{Observation Level Holdouts}
\label{s:eval:performance:basic_hold}
We begin with a basic evaluation of our model performance using a 10\% sample of our dataset as unseen data; we randomly select observations for inclusion in this unseen test set from our full set of labelled observations.

\begin{figure*}[t]  %
  \centering
  \begin{subfigure}[t]{0.32\textwidth}  %
    \includegraphics[width=\linewidth]{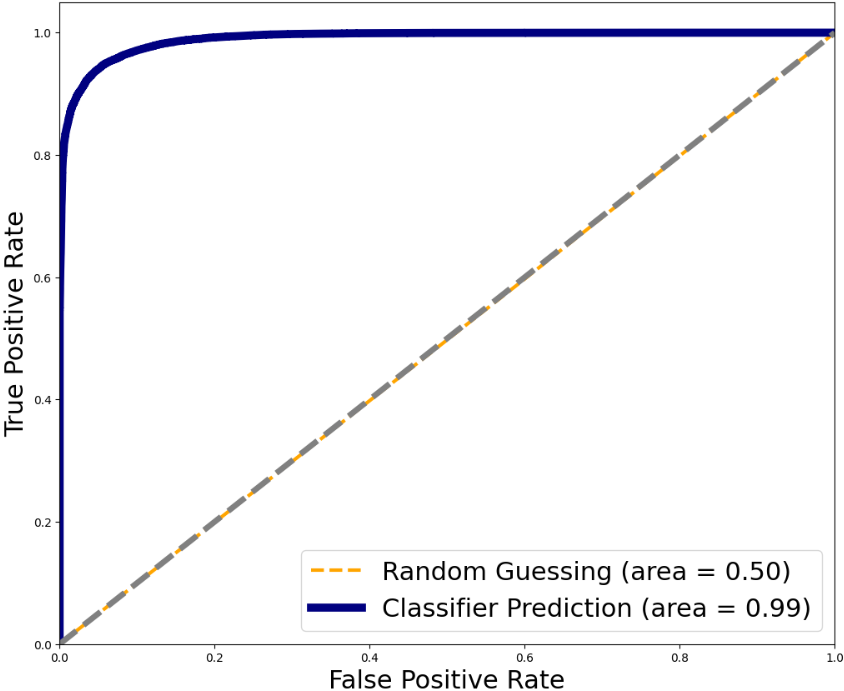}  %
    \caption{ROC curve for basic hold-out observations (unseen/test set).}
    \label{fig:roc_basic}
  \end{subfigure}%
  \hfill  %
  \begin{subfigure}[t]{0.32\textwidth}  %
    \includegraphics[width=\linewidth]{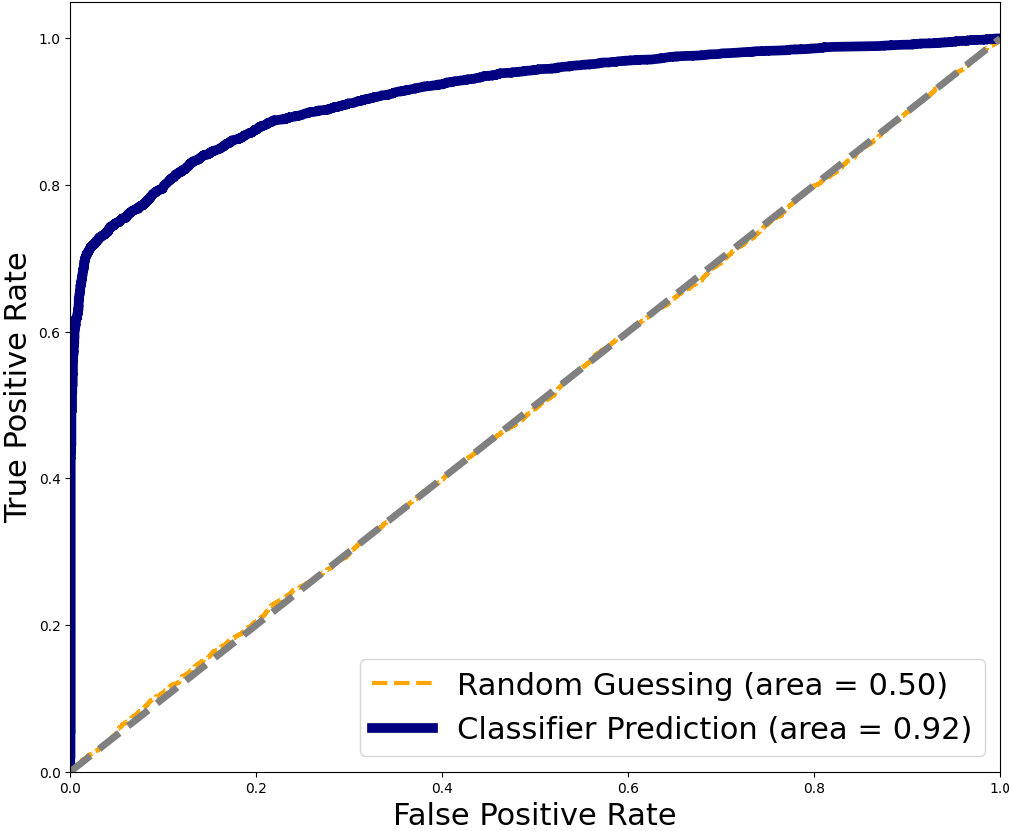}  %
    \caption{ROC curve for FCC adjudicated challenges on the unseen set.}
    \label{fig:roc_fcc_adj}
  \end{subfigure}%
  \hfill  %
  \begin{subfigure}[t]{0.32\textwidth}  %
    \includegraphics[width=\linewidth]{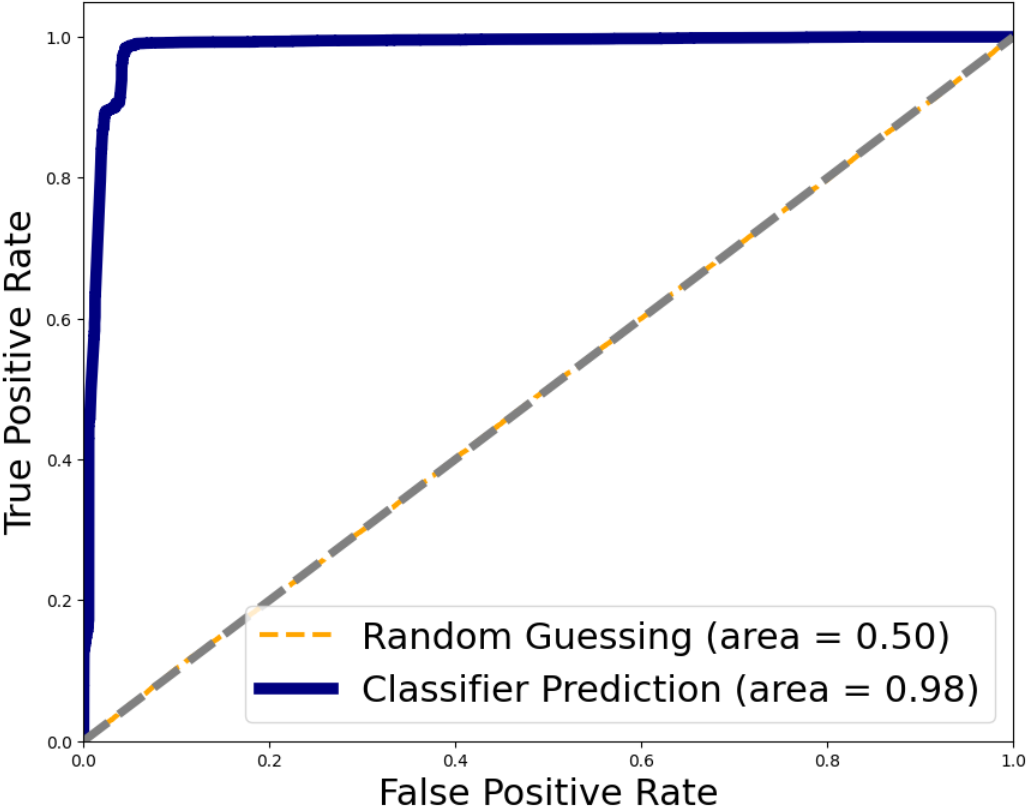}  %
    \caption{ROC curve for stratified evaluation of hold-out states (randomly selected).}
    \label{fig:roc_hold_states}
  \end{subfigure}
  \caption{The ROC scores for all three cases demonstrate the classifier's ability to distinguish between served and unserved cases, even in stratified holdout state observations (c). Although there is a slight decline in performance for the FCC-adjudicated challenges on the unseen set, the model's accuracy remains above 90\% (b).}
  \label{fig:roc_combined}
\end{figure*}

Model performance relative to the naive random guessing baseline is shown in Figure~\ref{fig:roc_basic}.
Our model performs very well, achieving an AUC of 0.99 and F1 score of 0.93, significantly outperforming the naive random guessing baseline.
The high ROC AUC score, close to 1, suggests that the model effectively discriminates between positive and negative instances in the test cases. 
In other words, our developed classifier correctly identifies most failed challenge observations.

We next restrict our hold out sample to only FCC adjudicated challenges.
FCC adjudicated challenges represent a relatively homogeneous subset of challenges: these challenges are resolved in a standardized manner by FCC staff after reviewing evidence provided by both challenger and provider.
Focusing on this set additionally controls for differences in provider responsiveness as well as challenger tenacity.
We define a new hold-out set of 10\% of all observations labelled only from these challenges and again evaluate model performance (Figure~\ref{fig:roc_fcc_adj}).

Here, our model achieves an AUC of 0.92, slightly underperforming against the baseline on labelled data from all sources.
Specifically, a precision of 0.78 for the negative class indicates that about five out of one of the 'unmodified' cases, which the FCC overturned (as shown in Table ~\ref{t:challenge_outcomes}), were incorrectly identified by our model, in reality, they were 'modified' (FCC Upheld in Table ~\ref{t:challenge_outcomes}). This suggests a slightly cautious approach by the model in selecting positive classes for FCC adjudication, possibly due to the limited amount of evaluation data (only 11k cases as support for evaluation).

However, the performance is still better than random guessing. With a combined F1 score of approximately 0.84 and an ROC AUC of 0.92, the model demonstrates predictive ability, indicating that it can effectively distinguish between classes under constrained data conditions. 

\subsubsection{Stratified Evaluation of Holdout States}
\label{s:eval:states}
Next, we randomly selected around 10\% of the total 56 states and territories as stratified unseen states for assessment. 
This means that any filling observations from these randomly selected states were not included in the training or validation sets, and this holdout set was used for performance evaluation.
We focus on this scenario for multiple reasons.
First, given the significant degree of state-to-state variation in participation in the challenge process, we wanted to understand if our model could generalize across states with varying degrees of challenges.
Second, given the underlying BSL Fabric changes with each release, we wanted to understand how well our model could perform on previously unseen geographies.
We randomly selected Nebraska (NE), Georgia (GA), Oklahoma (OK), Missouri (MO), Indiana (IN), and South Carolina (SC) as holdouts; we note that the state of Nebraska (NE) faced highest most location challenges in the initial BDC fillings (Figure~\ref{fig:state_wise_challenge}).

Overall, our model effectively distinguishes 'unserved' cases on unseen state-level data, achieving an ROC AUC of 0.98 (Figure~\ref{fig:roc_hold_states}). This indicates that the model maintains a high level of performance even when applied to completely new states.

\subsubsection{Classifier performance on major ISP's fillings}
\label{s:eval:states:major_isp}
We next evaluated how well our model performed in classifying major ISPs in each of these held-out states.

\begin{figure*}[t]  %
  \centering
  \begin{subfigure}[t]{0.5\textwidth}  %
    \includegraphics[width=\linewidth]{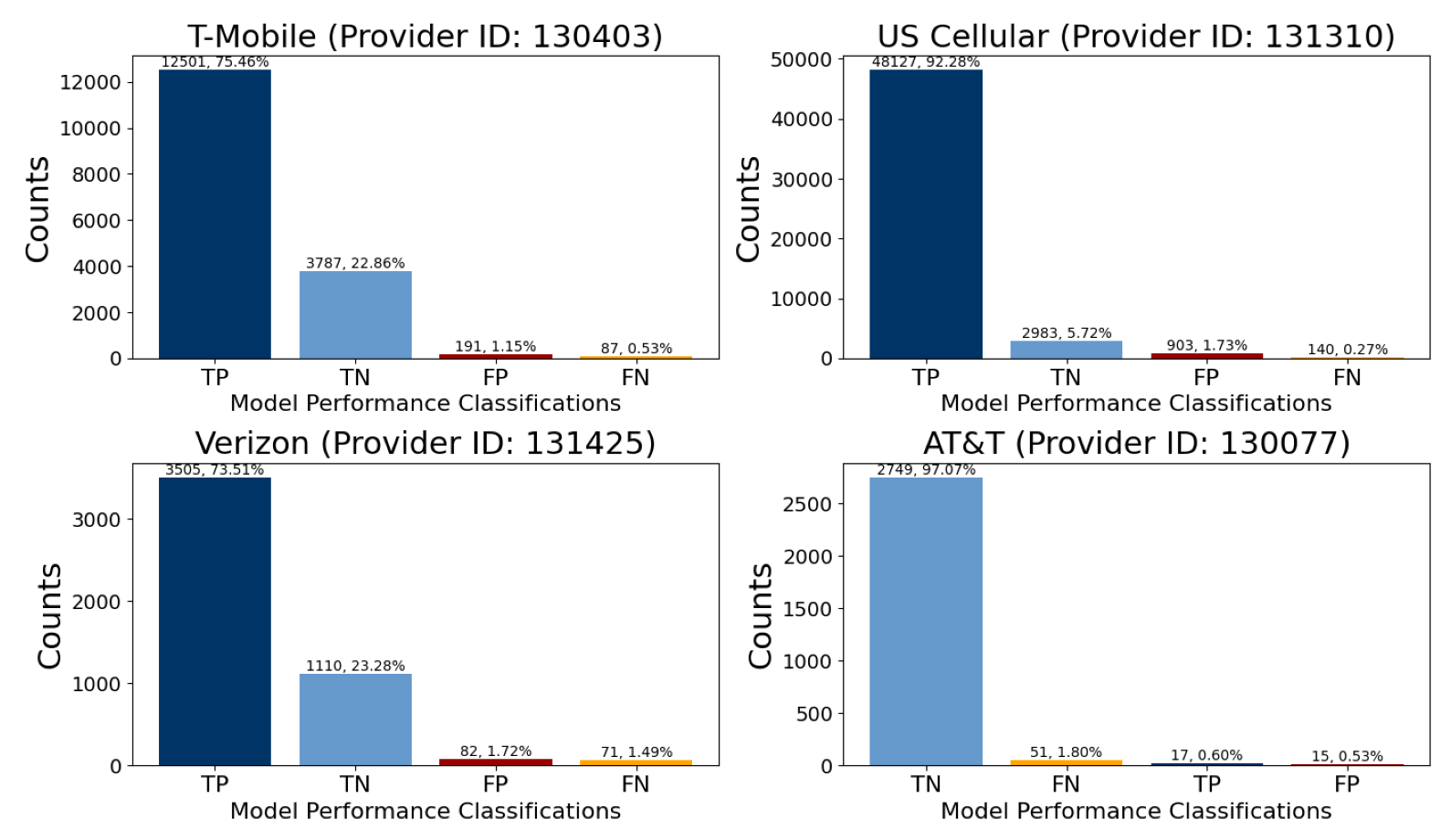}  %
    \label{fig:top_4_isp}
  \end{subfigure}%
  \hfill  %
  \begin{subfigure}[t]{0.5\textwidth}  %
    \includegraphics[width=\linewidth]{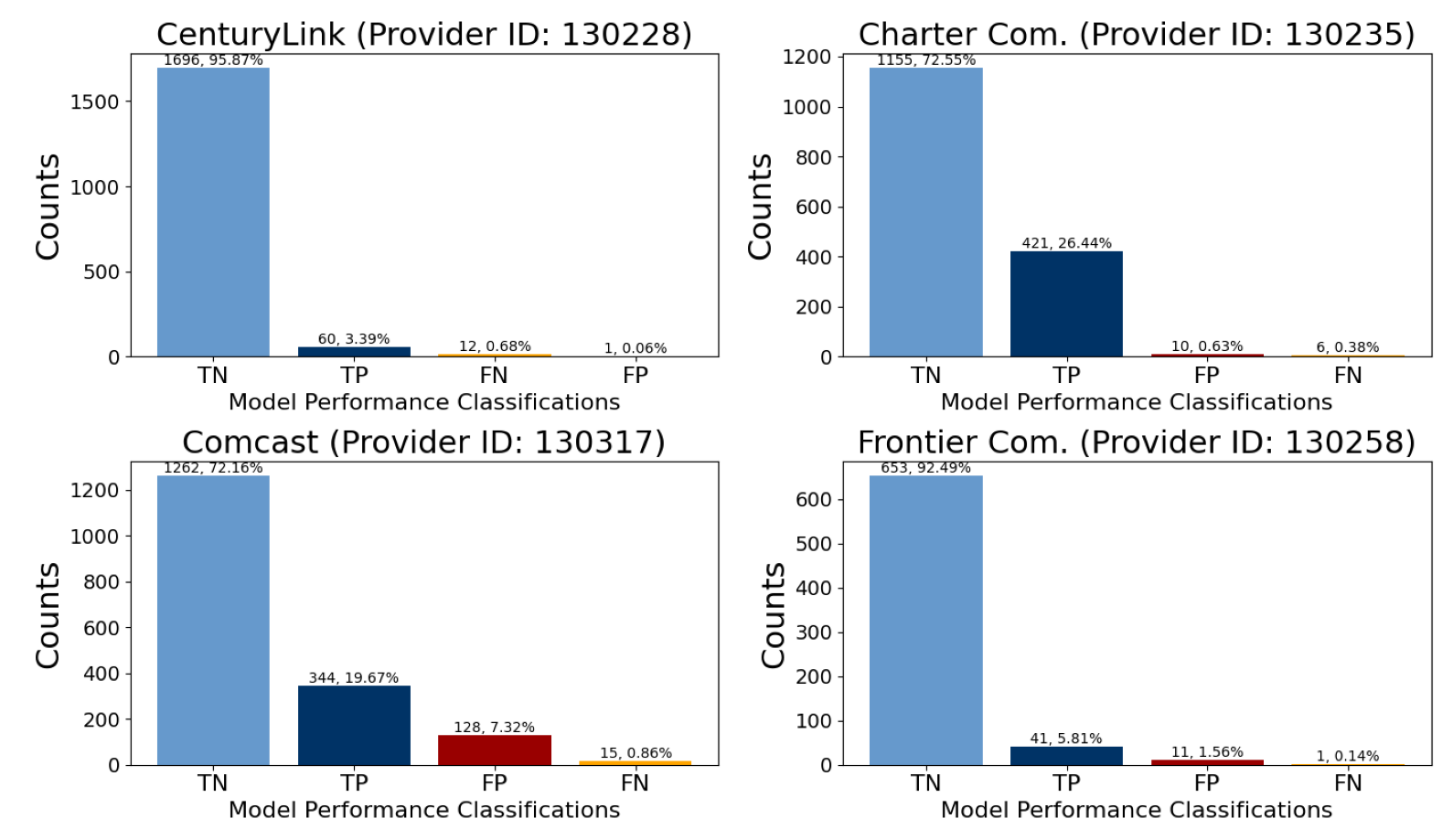}  %
    \label{fig:top_8_isp}
  \end{subfigure}
  \vspace{-0.2 in}
  \caption{Distribution of classifier prediction accuracy metrics for the major eight ISPs in holdout states. This indicates that our model consistently performs better (higher true case rates) for ISPs with relatively higher numbers of observations, such as T-Mobile and US Cellular.}
  \label{fig:top_isp_performance}
\end{figure*}

We identified eight of the largest terrestrial ISPs by locations served throughout the US.
Figure ~\ref{fig:top_isp_performance} shows a high rate of true positive and true negative cases. A true positive occurs when the model correctly predicts a "modified" case, and a true negative occurs when the model correctly predicts an "unmodified" case. This indicates that our model effectively distinguishes between both classes in the majority of cases.

From the figure, we can see that around 7\% of Comcast cases are false positives, meaning our model predicts something suspicious on those hexes. There could be three major reasons for this: no one challenged those suspicious cases, Comcast successfully contested those challenges, or our model is over-predicting against their filings. False negatives, while relatively low in percentage, indicate the model's shortcomings. In other words, the model failed to capture those (false negative) few "modified" or "unserved" instances.

\subsubsection{Dataset Utility}
Finally, we extend our stratified state holdout strategy to evaluate our approach for augmenting the original public challenge dataset with private changes and crowdsourced speed test-derived likely served locations.
To do this, we retrain our model on four subsets of our labelled dataset based on the source of each label: only public challenges, public challenges and private changes, public challenges plus likely served locations, and the full labelled dataset.
We would expect to see improved performance on the augmented balanced dataset with likely served locations (Synthetic).

\label{s:eval:states:aug_imp}

\begin{figure}
    \centering
    \includegraphics[width=0.85\linewidth]{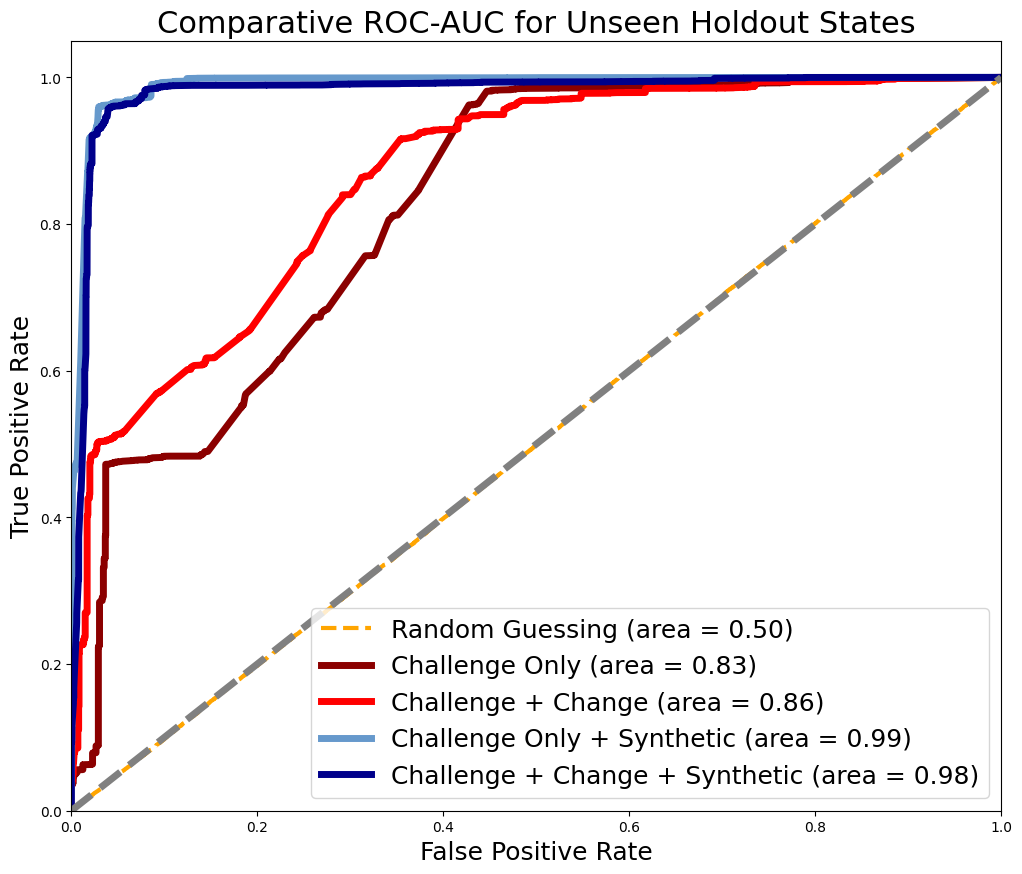}
    \caption{The comparative analysis of model performance metric (ROC) on held out states,  shows that performance improves as we incrementally add different labels, particularly after incorporating synthetic labelings. We selected the final dataset, which combines challenge, change, and synthetic labels, because it enables us to label our dataset across a wide range of providers and geographic areas.}
    \label{fig:comparative_roc_vf}
\end{figure}

The comparative analysis of ROC-AUC scores (Figure ~\ref{fig:comparative_roc_vf}) demonstrates the enhanced performance of the classifier when utilizing settings that include Challenges, Changes, and likely served location data. Also, Focusing on the F1 score, the initial model exhibited aggressive behavior (tending to predict locations to be unserved) when trained solely with challenge data or a combination of challenge and change data. 
This was due to these data sources emphasizing locations removed from providers' filings, which in turn led to a significant class imbalance and thus the bias towards predicting locations to be unserved.
By incorporating our synthetic likely served locations, we significantly improve the model performance.
Similarly, the ROC-AUC score increased and became close to 1 when considering the combined effect of challenge data, change data, and synthetic labels. This improvement underscores the efficacy of data augmentation in enhancing the model's capability to predict beyond mere challenge scenarios.

\subsection{Case Study: Jefferson County Cable}
We lastly turn to one case in which we do in fact have ground truth about a provider's service availability, Jefferson County Cable (JCC).
This provider intentionally overstated their service area in their initial \bdc filing to prevent an unserved area in which they intended to expand from being eligible for BEAD funding~\cite{brodkin2024jefferson}.
JCC was fined by the FCC for this intentionally inaccurate filing~\cite{fcc2024jefferson}.
Comparing the initial area that JCC reported as served with their later reported area, we observe that a collection of locations on the western side of their service area was subsequently removed.

\begin{figure}
    \centering
    \includegraphics[width=.9\linewidth]{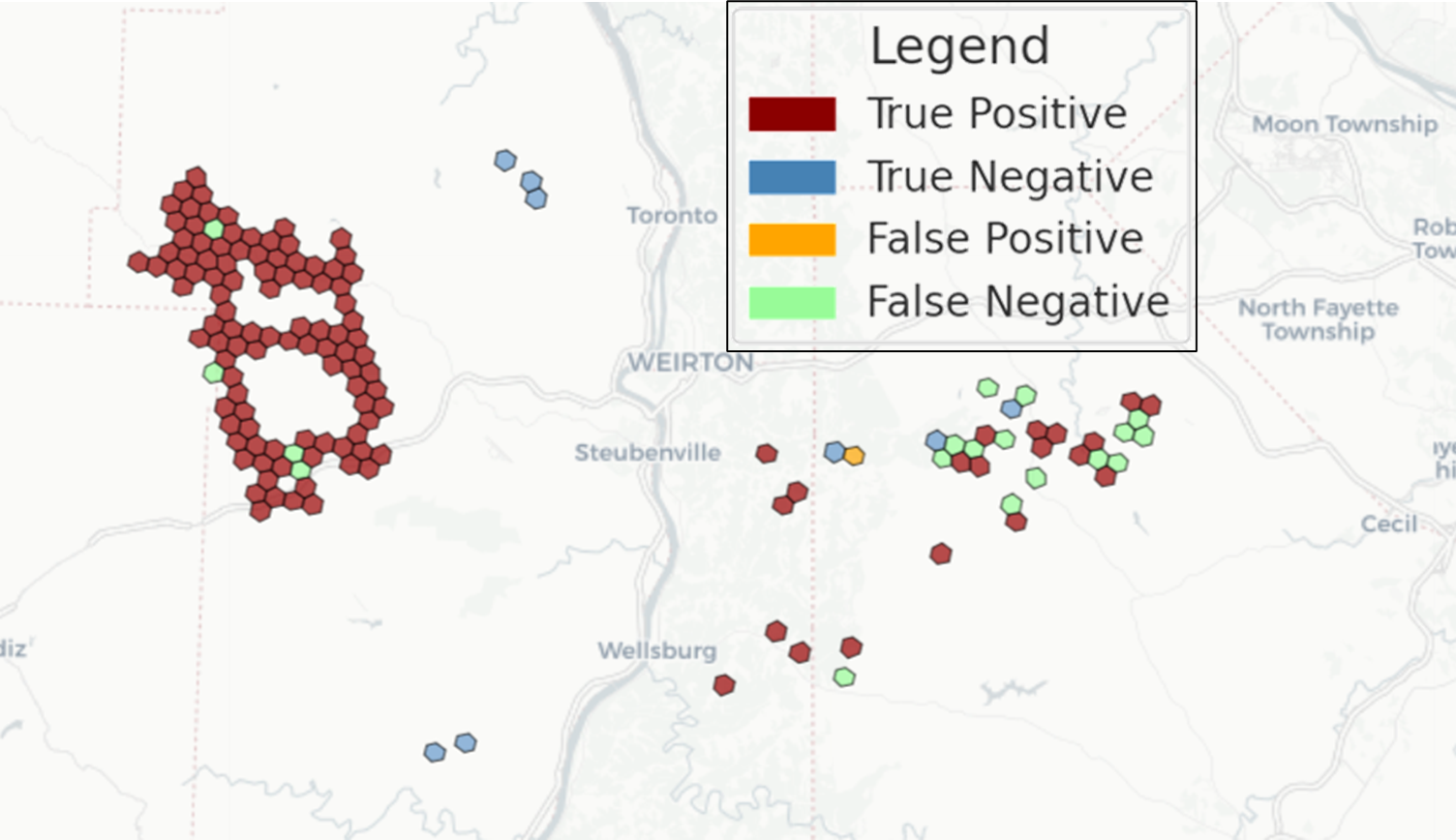}
    \caption{Model outputs on Jefferson County Cable, considering only labelled data. Positive indicates our model predicts the cell to be \emph{unserved} by this provider. Our model identifies the red region in the west where this provider falsely claimed to provide service.}
    \label{fig:jcc_map}
\end{figure}

Our model is able to identify the section of JCC's initial filing that corresponds to their misrepresentation.
We trained using a similar state holdout strategy as described in Section~\ref{s:eval:states} to remove all states bordering the service area of JCC, ensuring JCC and nearby challenges were not including in our training set.
Figure~\ref{fig:jcc_map} demonstrates that our model accurately classifies most of their false service claims as likely incorrect, while classifying their actual service area as largely likely served.

The case of JCC represents an important example of why systems like ours are useful for validating provider claims.
Although JCC had failed individual challenges, the sequence of events that led to enforcement action from the FCC only began after JCC unintentionally admitted in email to a competitor that they had misrepresented their service claims in the \bdc process.
Without this (un)lucky break, it is not clear that JCC's misreported claims would have been fully removed from the \nbm, thus rendering otherwise unserved locations ineligible for funding.
We argue tools like ours can assist policymakers in identifying areas that are likely overreported and further assist challengers in targeting limited resources to in areas where challenges are likely to succeed.

\section{\MakeUppercase{Conclusion and Discussion}} 
\label{s:discussion}
In this work, we have developed a novel dataset of broadband availability leveraging a combination of regulatory filings from the \nbmFull along with crowdsourced speed tests.
To our knowledge, this is the first work that directly associates self-reported provider service claims with measured network data.
We use this dataset to train a model to classify locations as likely served or unserved, enabling us to determine which portions of a provider's service claims are likely to fail challenges.
Our model performs effectively on a variety of unseen data scenarios, consistently achieving ROC AUCs above 0.9 and successfully identifying a well-publicized example of an intentionally misrepresented portion of a \bdc filing.
Despite strong performance in our benchmarks, our approach faces a number of limitations, and we consider this work an initial attempt to automatically identify problematic provider filings in the \bdc process.

\textbf{Limitations.}
We do not evaluate our dataset against actual ground truth.
Efforts like Measuring Broadband America~\cite{burger2023measuring} provide known locations with service at the level of an individual provider, and in future work we will evaluate the correspondence between our likely served locations and this data.
Nevertheless, MBA only covers a small set of providers; ground truth validation remains an ongoing challenge for the community.
Moreover, prior work~\cite{saxon2022we,lee2023analyzing} demonstrates that the underlying crowdsourced speed test data we use under-represents coverage in census tracts with higher proportions of low-income and non-white populations; we assume that our likely served locations similarly under-represent these areas. Our results thus likely overemphasize providers serving these areas, though given recent results on "digital redlining"~\cite{paul2023decoding,YinSankin2022} suggest that these same populations are subject to inequitable coverage and pricing, additional scrutiny of provider claims in these areas may be warranted.

Our approach is heavily dependent on an effective challenge process to help generate validated reports, and it is not clear how well our approach will generalize to future releases of the \nbm. Moreover, over time shifts in the underlying Fabric dataset as well as changes in ISP behavior may reduce the performance of our model.
It remains to be seen how well our change-based dataset will augment this lack of challenges.
Validating our model's performance across releases is an area of future work.

Finally, in this work, we have only considered the presence or absence of service from a provider, the most common cause of challenges to the \nbm.
Nearly as common (Table~\ref{t:reasons_for_challenges}) are scenarios where a provider overstates the level of service they can provide at a location; our technique does not capture this.
We leave this investigation of this to future work.

\textbf{Applying our results.}
One obvious application of our results would be to identify specific ISPs or locations where our model suggests a mismatch between data reported in the \nbm and our predictions.
We consciously choose not to report this in this work as \emph{this is not the intended use of our model}, and we are conscious of unintended data cascades~\cite{sambasivan2021everyone} resulting from misuse.
Our model is intended to help organizations preparing challenges to more effectively prioritize challenges; the regulatory process of formal challenges protects providers, communities, and policymakers by ensuring a consistent standard applies to challenges and thus the overall integrity of the \nbm.

\appendix

\section*{\MakeUppercase{Note}}
This is a preprint of a paper that has been accepted to the ACM Internet Measurement Conference (IMC) 2024, to be held in Madrid, Spain, from November 4–6, 2024. The final version will be published and available in the ACM Digital Library.

\section*{\MakeUppercase{Acknowledgments}}
We would like to thank the anonymous reviewers for
their valuable feedback that improved our paper, as well as Wesley Woo for many discussions that helped refine our approach during the development of this work.
We also thank Robert Martin, Nick Pappin, Christine Parker, Alexis Schrubbe, Michael Wasser, and Sascha Meinrath for their thoughtful discussions on working with FCC BDC data.

\renewcommand{\refname}{\MakeUppercase{References}}

\bibliographystyle{ACM-Reference-Format}

\section*{APPENDIX}
\section{Ethics}
\label{s:appendix:Ethics}
Our use of data in this paper raises no ethical concerns. All data used in this paper is from public sources including \nbmFull, Ookla Speedtests, and MLab data, and does not use any data from CostQuest; this was a conscious choice to ensure there are no restrictions on downstream use of our results. The only exception to this is the analysis of BSLs missing in the public National Broadband Map data discussed in Section \ref{s:datasets:location}. 

The models developed in this work can be used to characterize the validity of regulatory filings of ISPs.
We consider the output of our models to be suggestive, rather than definitive, and we do not recommend their use for regulatory enforcement.

\section{Terminology}
\label{s:appendix:terminology}

The terminology and definitions used in this work are illustrated in Table ~\ref{tab:terminology_definitions}.

\begin{table*}[h!]
\centering
\begin{tabular}{|p{3.5cm}|p{12cm}|}  %
\hline
\multicolumn{1}{|c|}{\textbf{Term}} & \multicolumn{1}{c|}{\textbf{Definitions}} \\ \hline
Failing a challenge & A "Fixed Challenge" is a process that allows individuals and organizations to dispute an ISP's availability claims. Failing a challenge means the ISP's claim is disproven, indicating that the provider's assertion was inaccurate. \\ \hline
True Positive (TP) & A positive outcome generally means our model predicts a challenge failure or identifies a suspicious claim by an ISP. A True Positive occurs when our model correctly predicts a challenge failure or identifies an unserved case. \\ \hline
False Positive (FP) & This occurs when our model predicts a challenge failure or identifies a suspicious ISP claim, but the claim is actually valid. This indicates the model's aggressiveness. \\ \hline
True Negative (TN) & In contrast to positive cases, a negative outcome means that the challenge will not fail, indicating that the ISP's original claim is valid. A True Negative occurs when our model correctly identifies valid claims. \\ \hline
False Negative (FN) & This occurs when the model predicts an ISP claim as valid, but the claim is actually a failure of challenges or an unserved case that the model failed to detect. False Negatives represent the actual shortcomings of our model. \\ \hline
\end{tabular}
\caption{Terminology and definitions}
\label{tab:terminology_definitions}
\end{table*}

\section{Mapping ASN to Provider ID}
\label{s:appendix:asn_mapping_methodology}
\textbf{Step 1: Map ASN to POC.}
We first map ASNs to their corresponding POCs. This involves analyzing the relationships between ASNs, Organizations (ORGs), Networks (NETs), and POCs as specified in the ARIN Bulk Whois data structure \cite{arin_bulk_whois}. We concluded the following three potential mappings from our analysis:
\begin{itemize}
    \item ASN $\rightarrow$ POC
    \item ASN $\rightarrow$ ORG $\rightarrow$ POC
    \item ASN $\rightarrow$ ORG $\rightarrow$ NET $\rightarrow$ POC
\end{itemize}

\textbf{Step 2: Prepare the Provider ID Data.}
We augment the BDC Provider ID table \cite{fcc_bdc_provider_id_table} with additional data from FRN registration.
This includes information such as email addresses, company names, and physical addresses. Enriching this data set enables more effective matching in the subsequent matching phase.

\textbf{Step 3: Matching ASN to Provider ID.}
We employ four independent matching methodologies email, physical address, company names, and contact email domain by performing the following steps to match ASN to Provider ID:
\begin{enumerate}
    \item \textbf{Canonicalization:} Standardize the data to improve match accuracy:
    \begin{itemize}
        \item \textbf{Email:} Remove trailing whitespace.
        \item \textbf{Contact Email Domain:} Remove whitespace and filter out domains that are publicly available for registration.
        \item \textbf{Company Name:} Remove trailing “inc” and “llc” and all non-alphanumerical and non-whitespace characters, convert to lowercase.
        \item \textbf{Physical Address:} Abbreviate words following USPS Publication 28 standard and remove all non alphanumerical and non-whitespace characters, convert to lowercase. 
    \end{itemize}
    \item \textbf{Prepare Contact Information to Provider ID Mapping:} Construct mappings from the standardized contact information to Provider IDs.
    \item \textbf{Matching:} For each ASN, match the standardized contact information against the mappings to determine the Provider ID.
\end{enumerate}
We define the set of ASNs associated with a provider to be the union of ASNs identified through each procedure.

\section{H3 Grid System}
\label{s:appendix:hex_system}

H3 is a discrete global grid system that defines a set of hexagonal tiles that cover a sphere, providing a way to tile the earth’s surface with cells of roughly equal size ~\cite{sahr2003geodesic, uber2018h3}. These tiles can be recursively subdivided into smaller hexagonal tiles, with the level of subdivision referred to as the “resolution” of the H3 grid. The finest publicly available spatial resolution for Broadband Serviceable Locations (BSLs) is at H3 resolution 8. Each H3 resolution 8 tile covers approximately 0.5 km², and around 690 million such tiles are required to cover the earth ~\cite{uber2018h3, h3geodocs2023}.

The FCC’s National Broadband Map provides a mapping between BSLs and their corresponding H3 hex tiles at resolution 8, and we directly use this spatial relationship in our project.

\begin{figure}
    \centering
    \includegraphics[width=1\linewidth]{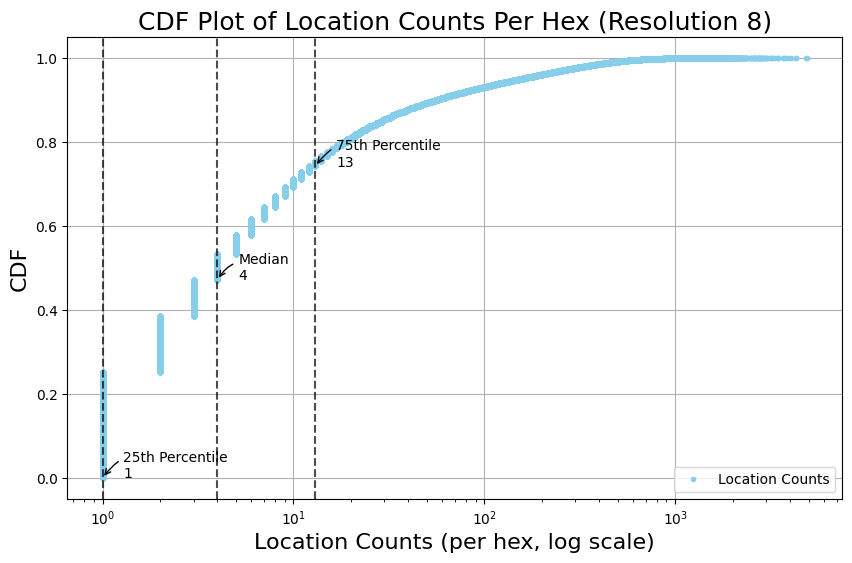}
    \caption{The median number of Broadband serviceable locations (BSL) per H3 resolution 8 cell is four.}
    \label{fig:h3_loc}
    \vspace{-0.1in}
\end{figure}

An individual H3 tile can encompass multiple locations; in this work, however, our fundamental unit of observation is a combination of <provider, technology, H3 resolution 8 grid>. When multiple Broadband Service Locations (BSLs) are reported within a single H3 tile for a given ISP, we adopt the maximum of their reported Maximum Download/Upload speeds at the BSL level. Additionally, if any BSL within a hexagon is successfully challenged, we consider the entire hexagon to be challenged. We used this approach because proving a challenge requires strong evidence, and the small size of an H3 tile at resolution 8 makes it practical to treat the entire tile as challenged if any part of it is.

\textbf{Reprojection of Ookla Dataset.} As previously mentioned, the public Ookla dataset is organized into 'quadkey' tiles~\cite{bing_maps_tile_system}, which we need to align with our other datasets using H3 tiles. To achieve this, we reproject the Ookla dataset from quadkeys to H3 grid tiles. Given that the spatial resolution of Ookla's data is typically finer than that of the H3 grid tiles, most quadkey tiles fit entirely within a single H3 tile. In cases where a quadkey tile spans multiple H3 grid tiles, we map it to each relevant H3 tile. We then aggregate Ookla's data into the H3 grid, summing the number of tests and devices observed per H3 tile, while recording the maximum reported average throughput and minimum observed latency.

\section{Interpretation of Model Results}
\label{s:appendix:interpretation}
We aim to interpret our tree-based XGBoost model and its results. This analysis is divided into two parts: In the first part, we will evaluate the model using SHAP analysis to understand feature importance and how these features drive the model's predictions. In the second part, we will characterize the model results by focusing on the key factors identified in the SHAP analysis.

\subsection{Factors Influencing Model Predictions}

To gain deeper insights into the factors driving our model’s performance, we focused on identifying the most contributory features and understanding the complex, non-linear relationships between these features and the target variable—whether a claim is classified as valid (served) or suspicious. We employed SHapley Additive exPlanations (SHAP) to accomplish this, as SHAP provides a comprehensive method for interpreting model predictions by assigning each feature a SHAP value that quantifies its contribution to the outcome \cite{lundberg2017unified}.

\begin{figure}
    \centering
    \includegraphics[width=1\linewidth]{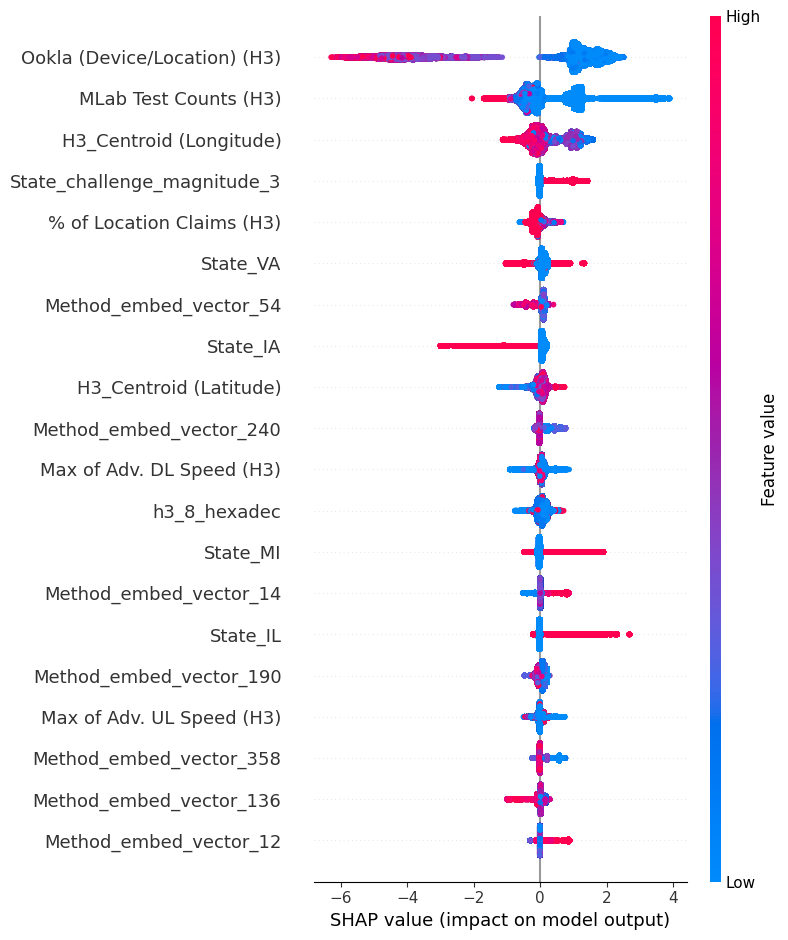}
    \caption{The SHAP summary plot reveals the top contributory features generally driving predictions used by the model to classify ISP filings as either valid or suspicious. Positive SHAP values indicate that a feature drives the prediction towards a suspicious claim, while negative values suggest that a feature leads the prediction towards a valid claim.}
    \label{fig:shap_summary}
\end{figure}

The SHAP summary plot (Fig.~\ref{fig:shap_summary}) visualizes the influence of features across the entire training dataset and highlights the top contributory features of our trained model as follows:

\paragraph{Ookla (Device/Location) and M-Lab Test Counts:} These two features emerged as the most influential in determining the model’s predictions. The SHAP summary plot (Fig.~\ref{fig:shap_summary}) indicates that higher values in “Ookla (Device/Location)” and “M-Lab Test Counts” generally drive the model towards predicting that a location is served (i.e., the claim is valid). Conversely, lower values of these features contribute to the model classifying a claim as suspicious (positive cases in terms of model outcome).

\paragraph{Location Information:} Features such as the H3 centroid data (longitude and latitude) and state-level data (e.g., State\_VA, State\_IA) also play a crucial role. The model leverages this geographical data to refine its predictions, with certain states and specific H3 centroid positions either increasing or decreasing the likelihood of a claim being suspicious.

For example, according to Fig.~\ref{fig:shap_summary}, higher values of H3 centroid longitude and lower values of H3 centroid latitude tend to shift the model's prediction towards the served class (indicating a valid, non-suspicious claim). In addition, when evaluating ISP filings from certain states, the model shows a tendency to predict claims as suspicious for filings from Illinois (State\_IL) and Michigan (State\_MI), in contrast to a lower likelihood of suspicion for filings from Iowa (State\_IA).

\paragraph{Other Features:} In addition to the primary contributory features, several other factors also influence the model’s predictions, either towards a positive (suspicious) or negative (valid) outcome. Notably, methodology embedding vectors—which represent the methodology explanations provided by ISPs during reporting—play a role in predicting whether a claim is suspicious or valid.

Another important feature is the percentage of BSL (Broadband Serviceable Locations) claims in a given H3 hexagon by an ISP, relative to the total number of BSLs in that area. This feature tends to lead the model towards valid predictions when the ratio is relatively higher, meaning that if an ISP claims the majority of BSLs within an H3 hexagon, the model is more likely to predict these claims as valid. When the ratio is lower, the model still considers this feature, but it may have less influence on the overall prediction, allowing the remaining features (such as maximum advertised downlink/uplink speeds) to play a more prominent role in the decision. 
Additionally, the features of maximum advertised downlink and uplink speeds reported by the ISP contribute to the model's predictions; however, the SHAP summary plot (Fig.~\ref{fig:shap_summary}) does not reveal a consistent trend in their impact. This indicates that while these speed features influence the predictions, their effect varies across specific observations (ISP, H3, Technology).

\begin{figure}
    \centering
    \includegraphics[width=1\linewidth]{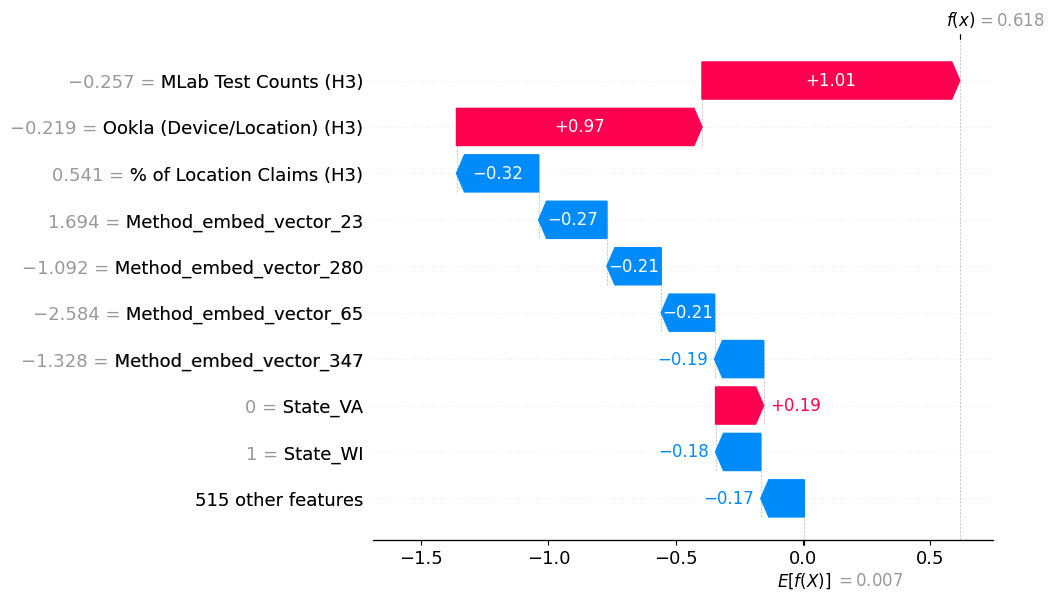}
    \caption{The SHAP waterfall plot illustrates the contributions, directions, and magnitudes of features for a single randomly selected prediction.}
    \label{fig:shap_waterfall}
    \vspace{-0.1in}
\end{figure}

Please note that in the summary plot, the direction of the model’s predictions is generalized across the entire training dataset, with each dot representing a single prediction. Individual predictions may not perfectly align with the general trends shown in the summary plot. To understand individual predictions more clearly, we randomly sampled single instances and visualized the contributory features for these particular predictions using the waterfall plot in Fig.~\ref{fig:shap_waterfall}. In the specific prediction shown, most of the other features had negative SHAP values that pulled the model's prediction towards the served case from the baseline. However, due to the higher magnitude of the positive SHAP values for the top two contributory features ("MLab Test Counts" and "Ookla Device/Location"), the sum of all feature contributions resulted in a final positive outcome (f(x) = 0.618 in Fig.~\ref{fig:shap_waterfall}). This means the model predicts that the ISP will fail in the challenge contest for that specific observation claim, indicating a suspicious claim.

\subsection{Characterization of Model Results}

Next, we focused on characterizing the model's results for the basic observation-level holdout (explained in Section \ref{s:eval:performance:basic_hold}) by examining key features identified through SHAP analysis. We analyzed the average values of prominent features for each class of model results, both by technology and state level.

Table ~\ref{tab:tech_wise_mapping} presents the classification report of model results for all five non-terrestrial technologies, revealing a common pattern. In each case, when the model correctly identifies valid ISP claims (True Negative cases), the average value of the Ookla indicator is greater than 1. This suggests that the model has learned that a high density of Ookla tests per location is indicative of a valid ISP claim (or served cases). Conversely, in True Positive (TP) cases, the TP class consistently shows the lowest average value of Ookla tests per location. This indicates that the model predicts suspicious (or unserved) cases when the Ookla indicator is at its lowest within that technology group. Similarly, a common pattern of the lowest MLab test counts is observed in most True Positive cases, suggesting that the model’s predictions are influenced by the 'MLab test counts' feature, with lower values associated with suspicious cases across different technologies.

\begin{table}[h!]
    \centering
    \begin{tabularx}{\columnwidth}{>{\centering\arraybackslash}X >{\centering\arraybackslash}X >{\raggedleft\arraybackslash}X >{\raggedleft\arraybackslash}X >{\raggedleft\arraybackslash}X}
    \toprule
    \textbf{Access Tech} & \textbf{Classify Report} & \textbf{Class \%} & \textbf{Ookla (Dev/Loc)} & \textbf{MLab Counts(K)} \\
    \midrule
    \multirow{4}{*}{LFW (71)} & TN & 17.27 & 2.33 & 30.50 \\ 
                              & TP & 78.37 & 0.16 & 3.95 \\ 
                              & FN & 1.69 & 0.55 & 26.53 \\ 
                              & FP & 2.67 & 0.48 & 12.70 \\ 
    \hdashline
    \multirow{4}{*}{ULFW (70)} & TN & 58.29 & 1.74 & 3.89 \\ 
                               & TP & 40.53 & 0.13 & 2.60 \\ 
                               & FN & 0.81 & 0.77 & 5.46 \\ 
                               & FP & 0.36 & 0.74 & 6.57 \\ 
    \hdashline
    \multirow{4}{*}{Fiber (50)} & TN & 42.61 & 1.70 & 109.21 \\ 
                                & TP & 51.92 & 0.20 & 21.93 \\ 
                                & FN & 4.68 & 0.32 & 144.52 \\ 
                                & FP & 0.78 & 0.34 & 57.24 \\ 
    \hdashline
    \multirow{4}{*}{Cable (40)} & TN & 34.90 & 1.91 & 153.07 \\ 
                                & TP & 54.41 & 0.11 & 68.62 \\ 
                                & FN & 7.67 & 0.25 & 81.40 \\ 
                                & FP & 3.03 & 0.19 & 113.34 \\ 
    \hdashline
    \multirow{4}{*}{Copper (10)} & TN & 83.25 & 1.58 & 33.26 \\ 
                                 & TP & 9.68 & 0.13 & 14.06 \\ 
                                 & FN & 6.47 & 0.19 & 19.62 \\ 
                                 & FP & 0.60 & 0.19 & 10.11 \\ 
    \bottomrule
    \end{tabularx}
    \caption{The table presents classification reports by access technology and the average top feature values by class. It clearly shows that for all technologies, an Ookla test density greater than 1 drives the model to predict valid claims correctly. In this context, LFW (71) stands for Licensed Fixed Wireless, and ULFW (70) refers to Unlicensed Fixed Wireless.}
    \label{tab:tech_wise_mapping}
\end{table}

Then we further characterized the model results at the state level. To do this, we randomly sampled 10 states from those where we have enough unseen holdout test sets (where "enough" means the state-level test set observation count is greater than or equal to 1,000). Table ~\ref{tab:state_predictions} indicates that there is variation in model accuracy across different geographic locations, such as states. For example, the model predictions in Idaho (ID) show perfect accuracy, as there are no false cases. However, states like Virginia (VA) and New York (NY) exhibit around 20\% inaccuracies in predictions, as indicated by the false positive (FP) and false negative (FN) rates.

For instance, in Idaho (ID), the average values for True Positive and True Negative cases are clearly distinguishable in the features represented in the table. In contrast, in New York (NY), the average values for True Positive and False Positive cases in the "Max Advertised Uplink Speed" and "Ookla Indicator" features are much closer, which confuses the model and leads to difficulty in distinguishing between served and unserved cases.

In Virginia (VA), none of the class averages for the "Ookla Indicator" feature exceed 1, which is unusual in this analysis, with the exception of Maine. The lack of Ookla speed tests in Virginia appears to contribute to a higher rate of false predictions by our model. Additionally, the average 'Max Advertised Downlink Speed' in Virginia shows relatively close values between True Negative and False Negative cases, which further confuses the model, leading it to incorrectly predict more served cases. In this context, a 17\% False Negative rate in Virginia means that, while the ISP's claim was originally disproven in challenges or identified as unserved, our model incorrectly predicts them as served or valid claims.

\begin{table*}[h!]
    \centering
    \begin{tabularx}{\textwidth}{>{\centering\arraybackslash}p{1.5cm} >{\centering\arraybackslash}X >{\raggedleft\arraybackslash}X >{\raggedleft\arraybackslash}X >{\raggedleft\arraybackslash}X >{\raggedleft\arraybackslash}X >{\raggedleft\arraybackslash}X}
    \toprule
    \textbf{State} & \textbf{Classification Report} & \textbf{Class \%} & \textbf{Ookla (Dev/Loc)} & \textbf{MLab Test Counts (K)} & \textbf{Max Adv. DL Speed (Mbps)} & \textbf{Max Adv. UL Speed (Mbps)} \\
    \midrule
    \multirow{4}{=}{Nebraska (NE)} & TN & 9.15 & 1.34 & 9.66 & 158.11 & 100.96 \\ 
                                   & TP & 90.18 & 0.11 & 0.55 & 114.71 & 93.03 \\ 
                                   & FN & 0.07 & 1.01 & 98.12 & 731.29 & 715.86 \\ 
                                   & FP & 0.60 & 0.10 & 0.00 & 25.00 & 3.00 \\ 
    \hdashline
    \multirow{2}{=}{Idaho (ID)} & TN & 33.10 & 1.52 & 15.25 & 199.03 & 115.10 \\ 
                                 & TP & 66.90 & 0.15 & 1.13 & 278.48 & 131.20 \\ 
    \hdashline
    \multirow{4}{=}{Kentucky (KY)} & TN & 50.52 & 1.66 & 18.31 & 404.08 & 219.97 \\ 
                                   & TP & 47.69 & 0.12 & 0.68 & 48.22 & 5.51 \\ 
                                   & FN & 1.03 & 0.65 & 10.61 & 48.33 & 8.53 \\ 
                                   & FP & 0.76 & 0.00 & 1.44 & 118.45 & 13.55 \\ 
    \hdashline
    \multirow{4}{=}{Virginia (VA)} & TN & 33.42 & 0.57 & 62.99 & 709.54 & 324.73 \\ 
                                    & TP & 45.50 & 0.10 & 38.18 & 340.08 & 51.70 \\ 
                                    & FN & 14.38 & 0.17 & 97.65 & 907.20 & 212.80 \\ 
                                    & FP & 6.70 & 0.13 & 61.40 & 250.45 & 33.10 \\ 
    \hdashline
    \multirow{4}{=}{North Carolina (NC)} & TN & 42.16 & 1.47 & 46.10 & 433.99 & 176.13 \\ 
                                           & TP & 54.64 & 0.11 & 0.66 & 56.88 & 12.99 \\ 
                                           & FN & 2.45 & 0.44 & 5.10 & 92.41 & 8.08 \\ 
                                           & FP & 0.74 & 0.34 & 2.92 & 76.75 & 11.60 \\ 
    \hdashline
    \multirow{4}{=}{Michigan (MI)} & TN & 30.55 & 1.40 & 25.78 & 1469.48 & 1120.46 \\ 
                                    & TP & 59.15 & 0.06 & 33.78 & 381.87 & 145.24 \\ 
                                    & FN & 1.46 & 0.66 & 39.87 & 992.18 & 440.96 \\ 
                                    & FP & 5.13 & 0.06 & 38.19 & 759.67 & 75.82 \\ 
    \hdashline
    \multirow{4}{=}{Maine (ME)} & TN & 63.40 & 0.51 & 4.22 & 496.73 & 194.43 \\ 
                                 & TP & 33.75 & 0.07 & 0.41 & 562.59 & 104.85 \\ 
                                 & FN & 2.09 & 0.17 & 0.14 & 503.96 & 200.80 \\ 
                                 & FP & 0.75 & 0.09 & 0.39 & 501.67 & 230.22 \\ 
    \hdashline
    \multirow{4}{=}{New York (NY)} & TN & 18.33 & 1.55 & 229.45 & 2663.22 & 2468.24 \\ 
                                     & TP & 79.10 & 0.12 & 26.27 & 618.72 & 40.09 \\ 
                                     & FN & 15.17 & 0.51 & 99.87 & 218.56 & 126.40 \\ 
                                     & FP & 3.47 & 0.07 & 48.01 & 946.94 & 29.08 \\ 
    \hdashline
    \multirow{4}{=}{Ohio (OH)} & TN & 50.50 & 1.93 & 53.40 & 164.75 & 61.82 \\ 
                                & TP & 36.65 & 0.07 & 29.01 & 24.04 & 6.17 \\ 
                                & FN & 9.79 & 0.77 & 20.14 & 76.38 & 44.26 \\ 
                                & FP & 3.06 & 0.19 & 14.74 & 137.30 & 117.59 \\ 
    \hdashline
    \multirow{4}{=}{Florida (FL)} & TN & 67.40 & 4.01 & 35.40 & 403.32 & 261.43 \\ 
                                   & TP & 13.96 & 0.10 & 48.64 & 899.52 & 92.98 \\ 
                                   & FN & 5.16 & 0.17 & 22.21 & 1846.21 & 1503.75 \\ 
                                   & FP & 5.13 & 0.10 & 80.97 & 759.67 & 75.82 \\ 
    \bottomrule
    \end{tabularx}
    \caption{The table presents randomly selected state-wise classification reports along with the average values of prominent features by class. It shows that the model's accuracy varies significantly across states, ranging from perfect accuracy to deviations of over 20\%. The model also appears to learn differently for various features depending on the state. However, Ookla test density remains the primary driver of the model's results.}
    \label{tab:state_predictions}
\end{table*}

In summary, our additional analysis for explaining the model outcomes is as follows:

\begin{itemize}
    \item The SHAP analysis identifies the most important features and how they influence the model's predictions. Features such as Ookla test density, MLab test counts, and geographic location information are the primary contributors to our model’s predictions. Generally, higher values of third-party speed tests help the model learn to predict valid ISP claims. Additionally, features like advertised speed (UL/DL), the ratio of locations covered in a hex, and word embedding vectors from methodology explanations also contribute to the model’s results.

    \item The accuracy of the model result varies from state to state. Result accuracy decreases in regions with a lack of speed test data (e.g., Virginia). Furthermore, the dynamics of maximum advertised speed influence model predictions differently across states. In the absence of clear distinctions in these values, the model struggles to accurately predict outcomes (e.g., Virginia, New York).
\end{itemize}

\end{document}